\documentclass{sig-alternate-05-2015}
\usepackage{booktabs} % For formal tables
\usepackage{url}
\usepackage{latexsym}
\usepackage{amsmath}
\usepackage{algorithm}
\usepackage{graphicx}
\usepackage{subfigure}
\usepackage{csquotes}
\usepackage[export]{adjustbox}
\usepackage{tabularx}
\usepackage{enumitem}
\usepackage{tikz}
\usepackage{tikz-cd}
\usepackage{array}
\newcolumntype{M}[1]{>{\centering\arraybackslash}m{#1}}
\newcolumntype{N}{@{}m{0pt}@{}}

\begin{document}
\title{WikiM: Metapaths based Wikification of Scientific Abstracts}
%\titlenote{Produces the permission block, and
%  copyright information}
%\subtitle{Extended Abstract}
%\subtitlenote{The full version of the author's guide is available as
%  \texttt{acmart.pdf} document}

\numberofauthors{4} %  in this sample file, there are a *total*
% of EIGHT authors. SIX appear on the 'first-page' (for formatting
% reasons) and the remaining two appear in the \additionalauthors section.
%
 \author{
% % You can go ahead and credit any number of authors here,
% % e.g. one 'row of three' or two rows (consisting of one row of three
% % and a second row of one, two or three).
% %
% % The command \alignauthor (no curly braces needed) should
% % precede each author name, affiliation/snail-mail address and
% % e-mail address. Additionally, tag each line of
% % affiliation/address with \affaddr, and tag the
% % e-mail address with \email.
% %
\alignauthor
Abhik Jana\\
       \affaddr{Dept. of CSE}\\
       \affaddr{IIT Kharagpur}\\
       \affaddr{West Bengal, India -- 721302}\\
       \email{abhik.jana@iitkgp.ac.in}
 %2nd. author
\alignauthor
Sruthi Mooriyath\\
       \affaddr{SAP Labs India Pvt Ltd}\\
       \affaddr{India}\\
       %\affaddr{34014 Trieste, Italy}\\
       \email{sruthi.m01@sap.com}
 %3rd. author
\and 
\alignauthor Animesh Mukherjee\\
       \affaddr{Dept. of CSE}\\
       \affaddr{IIT Kharagpur}\\
       \affaddr{West Bengal, India -- 721302}\\
       \email{animeshm@cse. \\
       iitkgp.ernet.in}
  % use '\and' if you need 'another row' of author names
% 4th. author
\alignauthor Pawan Goyal\\
       \affaddr{Dept. of CSE}\\
       \affaddr{IIT Kharagpur}\\
       \affaddr{West Bengal, India -- 721302}\\
       \email{pawang.iitk@gmail.com}
 }

\if{0}
\numberofauthors{4} %  in this sample file, there are a *total*
% of EIGHT authors. SIX appear on the 'first-page' (for formatting
% reasons) and the remaining two appear in the \additionalauthors section.
%
 \author{
% % You can go ahead and credit any number of authors here,
% % e.g. one 'row of three' or two rows (consisting of one row of three
% % and a second row of one, two or three).
% %
% % The command \alignauthor (no curly braces needed) should
% % precede each author name, affiliation/snail-mail address and
% % e-mail address. Additionally, tag each line of
% % affiliation/address with \affaddr, and tag the
% % e-mail address with \email.
% %
\alignauthor 
Abhik Jana\\
       \affaddr{Dept. of CSE}\\
       \affaddr{IIT Kharagpur}\\
       \affaddr{West Bengal, India -- 721302}\\
       \email{abhik.jana@iitkgp.ac.in}
 %2nd. author
\alignauthor 
Sruthi Mooriyath\\
       \affaddr{SAP Labs India Pvt Ltd}\\
       \affaddr{India}\\
       \email{sruthi.m01@sap.com}
 %3rd. author
  % use '\and' if you need 'another row' of author names
% 4th. author
\and
\alignauthor 
Animesh Mukherjee\\
       \affaddr{Dept. of CSE}\\
       \affaddr{IIT Kharagpur}\\
       \affaddr{West Bengal, India -- 721302}\\
       \email{animeshm@cse.iitkgp.ernet.in}

\alignauthor 
Pawan Goyal\\
       \affaddr{Dept. of CSE}\\
       \affaddr{IIT Kharagpur}\\
       \affaddr{West Bengal, India -- 721302}\\
       \email{pawang.iitk@gmail.com}
  
}

\fi
% The default list of authors is too long for headers}
%\renewcommand{\shortauthors}{Jana et al.}

\maketitle
\begin{abstract}
In order to disseminate the exponential extent of knowledge being produced in the form of scientific publications, it would be best to design mechanisms that connect it with already existing rich repository of concepts -- the Wikipedia. Not only does it make scientific reading simple and easy (by connecting the involved concepts used in the scientific articles to their Wikipedia explanations) but also improves the overall quality of the article. In this paper, we present a novel {\em metapath} based method, {\bf WikiM}, to efficiently wikify scientific abstracts -- a topic that has been rarely investigated in the literature. One of the prime motivations for this work comes from the observation that, wikified abstracts of scientific documents help a reader to decide better, in comparison to the plain abstracts, whether (s)he would be interested to read the full article. We perform mention extraction mostly through traditional tf-idf measures coupled with a set of smart filters. The entity linking heavily leverages on the rich citation and author publication networks. Our observation is that various {\em metapaths} defined over these networks can significantly enhance the overall performance of the system. For mention extraction and entity linking, we outperform most of the competing state-of-the-art techniques by a large margin arriving at precision values of {\bf 72.42\%} and {\bf 73.8\%} respectively over a dataset from the ACL Anthology Network. In order to establish the robustness of our scheme, we wikify three other datasets and get precision values of {\bf 63.41\%-94.03\% } and {\bf 67.67\%-73.29\%} respectively for the mention extraction and the entity linking phase. 
 
\end{abstract}

% We no longer use \terms command
%\terms{Theory}

\keywords{Wikification, Scientific article, Mention Extraction, Entity linking, Metapath, Citation network, Author publication network}

%\maketitle

%\input{samplebody-conf}
\section{Introduction}

Wikipedia (introduced in 2001) is an online encyclopedia that has evolved as the largest repository of collaboratively curated encyclopedic knowledge having millions of articles, in more than 200 languages. %Each Wikipedia article hosts a rich collection of concepts along with a set of hyperlinks associating crucial terms to other wiki pages, which makes the article more meaningful and easy to follow. 
The reliable and refreshed knowledge base of Wikipedia makes it a very popular (Alexa Rank -- 7\footnote{http://www.alexa.com/siteinfo/www.wikipedia.org}) source of knowledge. Consequently, there is an increasing research drive to utilize this knowledge base for better interpretation of terms and expressions in a given text. Wikification, one such usage of Wikipedia introduced by Mihalcea and Csomai~\shortcite{mihalcea2007wikify}, is the process of identifying important phrases in a given text ({\em mention extraction}), and linking each of them to appropriate Wikipedia articles depending on their context of appearances ({\em entity linking}). \\

%is an entity linking task to identify the important terms in a given text ({\em mention extraction}) and then link these terms to the corresponding Wikipedia pages ({\em link disambiguation}). 

\noindent\textbf{What is available?} There has been a substantial amount of work in the literature which focuses on the entity disambiguation and linking task~\cite{ratinov2011local,davis2012named,sil2012linking,demartini2012zencrowd,wang2012targeted,alhelbawy2014graph,pershina2015personalized,pan2015unsupervised}. There has also been a lot of literature that contributes to the whole process of end-to-end linking. While some researchers focus on wikification of standard text~\cite{mcnamee2009overview,ji2010overview,ji2014overview}, there have been some efforts to wikify microblog text as well~\cite{genc2011discovering,cassidy2012analysis,guo2013link,huang2014collective}. Wikifier~\cite{cheng2013relational} is one such tool for wikification which adopts Integer Linear Programming (ILP) formulation of wikification that incorporates the entity-relation inference problem. Another attempt has been made by Yosef et al.~\shortcite{yosef2011aida}, where they propose a graph-based system AIDA, a framework and an online tool for entity detection and disambiguation. Moro et al.~\shortcite{Moro14entitylinking} presented Babelfy, a graph-based approach which exploits the semantic network structure for entity linking and word sense disambiguation. TagMe is also one of the software systems in this area presented by Paolo and  Ugo~\shortcite{ferragina2010tagme} which can annotate any short, poorly composed fragments of text achieving high on-the-fly accuracy. Thus, the research on entity linking targets on a broad range of text including newswire, spoken dialogues, blogs/microblogs and web documents in multiple languages. \\

\noindent\textbf{What is lacking?} Despite such a huge effort by the research community, there have been very few attempts to wikify scientific documents. Most of these studies are made specifically for the bio-medical documents. Some mention extraction tasks like human gene name normalization~\cite{hirschman2005overview,usami2011automatic,fang2006human,frisch2009litinspector}, discovery of scientific names from text~\cite{akella2012netineti} as well as the entity linking task in this domain~\cite{zheng2015entity} have been attempted by the researchers recently. Some efforts have been made in the geo-science literature~\cite{ma2015illuminate} as well. These approaches, however, are domain specific and do not go beyond mention detection. \\
\iffalse to the wikification aspect.

\fi %none of those tried to exploit the rich citation and collaboration network of scientific documents for link disambiguation

\noindent\textbf{Motivation:} Scientific articles are one of the most influential, authorized medium of communication among researchers allowing for appropriate dissemination of knowledge. %within the scientific community. 
To make this transmission effective, the terms or phrases used in a particular scientific article need to be understood by the readers without putting too much time or effort. % and spending much time. 
The need is further accelerated as the number of scientific articles are in millions, growing at a rate of approximately 3\% annually~\cite{bornmann2015growth}, and there is no well-accepted self-sufficient scientific vocabulary. It would be worthwhile to have an automated system which will do this task of entity linking on scientific articles. Abstract of a scientific article provides a glimpse of the whole article in a nutshell. It is a usual practice among the researchers to go through the abstract first to assess the importance and relevance of the article to their interest. Sometimes the availability of full scientific articles is also under certain terms and conditions (e.g., individual or institutional subscriptions), but the abstracts are publicly available. Thus researchers find it always better to perceive the main idea from the abstract before requesting access from the competent authority for the full article. Further, if a researcher wants to step into a new area of research, wikification could be very helpful to get acquainted with the new terminologies. In order to validate our hypothesis that wikified abstracts help in making better judgments as to whether to read the whole article, we conduct a survey where the wikified and the non-wikified versions of 50 random abstracts from different domains are shown to 10 researchers and 10 post-graduate students. Subsequently, they are asked to judge whether the wikified version of the abstract helped them in forming an overview of the scientific article, better than the non-wikified version. As per their responses, wikification was found to be helpful for 72\% cases; this shows the importance of wikification in this problem context. Further, when the researchers were given an article abstract from a domain different from their areas of research, they voted for the wikified version in 100\% cases, which straightaway supports our hypothesis. Therefore in this work, we focus on the wikification of scientific article abstracts which we believe should immensely help the readers to decide whether they would at all wish to access the full article (which might have a cost/subscription associated). 

We tried this wikification task on scientific abstracts using TagMe~\cite{ferragina2010tagme} that leads to poor results for mention extraction and entity linking with precision of 30.56\% and 58.91\% respectively. Similarly, other baselines like AIDA~\cite{yosef2011aida} and Wikifier~\cite{cheng2013relational} also do not perform well, achieving precision values of 8.54\%-23.88\% and 6\%-19.33\% respectively for mention extraction and entity linking tasks. This constitutes our motivation behind proposing a novel approach to wikify scientific articles that makes use of a variety of information specific to scientific texts. In the recent literature there has been burgeoning research where metapaths are closely studied in heterogeneous information network for several mining tasks like similarity search, relationship prediction, clustering etc.~\cite{Sun:2012:MHI:2371211,Sun:2013:PIM:2513092.2500492,Liu:2014:MRP:2661829.2661965}. Metapath based similarity~\cite{Sun11pathsim:meta} forms a common base for any network-based similarity that captures the semantics of peer similarity. Motivated by this stream of literature, we attempt to apply metapath based approach for entity linking which is a natural choice in this case as the citation and author publication networks are perfect sources for construction of meaningful and relevant metapaths. \\     
%\vspace{3mm}

\noindent\textbf{How we contribute?} The main contributions of this paper are summarized as follows:
\begin{itemize}[noitemsep,nolistsep,leftmargin=*]
\itemsep0em 
\item This is the first attempt to wikify scientific articles, not specific to any domain. Our method -- WikiM -- can wikify any scientific article abstract for which we have the underlying citation network, author publication network etc. 
%\item A collective inference approach is designed to address the linking problem through maximizing the agreement between the text of the mention document and the context of the entities of the knowledge base. 
\item We exploit metapaths between scientific document collections to address the entity linking problem. To extract different types of metapaths, we use the citation network and the author publication network.
\item We perform extensive evaluations over 3 different datasets.
%along with a benchmark dataset from SemEval. 
We find that the proposed system is able to improve upon the existing baselines and also gives good performance across various datasets consistently. We outperform most of the competing state-of-the-art techniques by a large margin arriving at precision values of {\bf 72.42\%} and {\bf 73.80\%} respectively for mention extraction and entity linking phase over a gold Standard dataset from the ACL Anthology Network. In order to establish the robustness and consistency of our scheme, we wikify three other datasets and get precision values of {\bf 63.41\%-94.03\% } and {\bf 67.67\%-73.29\%} respectively for the mention extraction and the link disambiguation phase. 
%\todo{Report all accuracy figures; you may also choose to report improvements.} 
We make available the code, the results of WikiM and other baselines as well as the gold standard annotations\footnote{https://tinyurl.com/ko84whm}.% here {}. {\color{red}{\bf Put anonymous link to the data and code.}}
\end{itemize}

\section{Methodology}
A two step approach is proposed for wikification of scientific documents. The mentions are the important phrases in the scientific document that are potential terms for wikification. As the first step, given a scientific abstract $ d $, we extract a set of important mentions $ M = \{m_{1}, m_{2},..,m_{n}\} $ as described in Section~\ref{sec:keyword}. In the second step as described in Section~\ref{ld}, for each mention $m$, we extract a list of candidate entities (Wikipedia links) $ C $ with surface form similar to $ m $ from Wikipedia and rank them according to the similarity scores calculated by metapath-based approach. Finally, we select the entity with the highest score as the appropriate entity for linking.
\iffalse
\begin{figure}
    \centering
    \includegraphics[width=0.5 \textwidth]{MTP_Overview.jpg}
    \caption{Approach Overview.}
    \label{fig:Approach Overview}
\end{figure}

\begin{algorithm}
\caption{Link\_Mentions}\label{euclid}
\begin{algorithmic}[1]
\Procedure{LinkMention($m_{i}$)}{}
%\State $\textit{stringlen} \gets \text{length of }\textit{string}$
%\BState \emph{top}:
\If {$|C| = \textit{1}$} \Return its wikipedia link
\Else 
\For {each $c_{i} \epsilon C$}
\State $N \gets \textit{all n-grams upto n = 3 of } c_{i}$ 
\State $s_{i} \gets N \cap M$
\EndFor
\State $max_{c_{i}} \gets \max(s_{i})$
\State $C' = {\{ c_{i} \colon s_{i} = max_{c_{i}}\}}$
\If {$|C'| > 1$}
%\State $CS_{i} \gets cos\_sim(d, WS_{m})$
\If {$conf\_score > 0.06$}
\State\Return Wikipedia link of $c_{i}$
\Else
\If {$CS(MP(d), WS_{m}) > 0.4$}
\State $l \gets \max(CS(d+$
\State $MP(d)), WS_{m})$
\State\Return Wikipedia link of 
\State $m_{i}$ corresponding to $l$
\EndIf
\EndIf
\EndIf
\EndIf
\EndProcedure
\end{algorithmic}
\end{algorithm}
\fi
%\vspace{1mm}
%\todo{Pawan: We can remove this schematic if we need space. This does not add much to the flesh of the paper. All that could be said is said in the text already.}
\subsection{Mention extraction} 
\label{sec:keyword}
%\vspace{1mm}
The first step in our approach is 
%to link identified mentions to the concepts in Wikipedia. Therefore, 
to identify important mentions from a scientific document abstract $d$. 
%we apply various publicly available natural language processing tools. 
After performing tokenization, % for most frequent words such as $the, a, of, for, in,$ etc., which often do not carry much meaning using NLTK toolkit~\cite{Loper:2002:NNL:1118108.1118117}.
text normalization is done specific to scientific documents to remove author names and year of publications; for example, Blei et al., 2003; Griffiths, 2004 etc. The next step is to apply POS tagging and identify a set $ F $ of textual fragments that contain sequences of nouns or adjectives of length up to three. %\todo{Please check this. Only 1 noun/adjective out of 3?}
Moreover, overlapping text fragments are handled by giving preference to the larger length fragment if it exists. For example, we do not recognize $ language $ and $ modeling $ separately, for $ language \text{ } modeling $.%, rather itself being a bi-gram seems to have greater significance in context of the sentence. %\todo{This is not clear. Gives meaning to the sentence ...?}

For each textual fragment $ f \in F $, we determine the linking validity corresponding to a decision as to whether it is matched with any Wikipedia concept. We have used python library `wikipedia' which wraps the MediaWiki API for this purpose. If there is no wikipedia article having similar surface form as the textual fragment we are looking for, the library function would return an error message. Only the fragments (a single- or multi-word expression) with positive validity values are taken as the candidate set of mentions $ M $. %While all the multi-word mentions are considered without any further processing, 
Then the set of single-word and multi-word mentions are ranked based on the tf-idf score.%\footnote{$tf-idf(m,d) = \frac{tf(m, d)}{max\{tf(i, d)\}}\log\frac{N}{df_m}$ where $N$ and $df_m$ are computed using the 2013 ACL Anthology Network.}.
%and predefined number of top ranked mentions are considered as potential keywords.  %\todo{Abhik, are you doing this in the algorithm?}

 %$ N $ is the total number of documents in the 2013 ACL Anthology Network and $df(m)$ is the total number of documents in which mention $m$ occurs, 

%For each abstract $d$, the number of mentions $|M|$ selected by the algorithm depends on the number of sentences in the abstract ($nS(d)$) as follows: $|M| = 4$ if $nS(d) <= 2$, $=8$ if $nS(d) <= 4$, $=12$ if $nS(d) > 4$. These numbers are chosen based on the manual inspection of a subset of scientific abstracts.
For each abstract $d$, the number of mentions $|M|$ selected by the algorithm depends on the number of sentences in the abstract ($nS(d)$) as follows:
\[|M| =  \left\{
	\begin{array} {lr}
	4 & nS(d) \leq 2 \\
    8 & 2 < nS(d) \leq 4\\
    12 & nS(d) > 4
	\end{array}
\right.
\]

These numbers are chosen after going through a few scientific abstracts manually.
%where $nS(d)$ gives the number of sentences in the abstract $d$.

%\todo{But do you rank bi-gram and trigram also in the same manner?}

%following two features:
%\begin{enumerate}
%\item $ TF(m, d) = \frac{count(m, d)}{max\{count(m^{'}, d)\}}$ to measure the raw frequency of $ m $ in document $ d $
%\item $ IDF(m)  = \log(\frac{N}{df(m)})$ 
%\todo{You need to tell the number of mentions being extracted by your algorithm.} 
%\end{enumerate}
%\vspace{-5mm}
\subsection{Entity linking}\label{ld}
The second step in our approach is to link the extracted mentions to appropriate Wikipedia entities. The detailed procedure of entity linking is described next.
\subsubsection{\bf{Candidate entity generation \& ranking}}\label{candidate_gen}
%\vspace{1mm}
For each mention $ m \in M$, identified in the previous step, we collect all the Wikipedia entities with surface forms similar to the mention's surface form, and consider them as candidates $ e \in E $ for the mention\footnote{We have used python library `wikipedia' which wraps MediaWiki API for this purpose}.
%\todo{Abhik, is this done via an API or how?} 

First, we assign a confidence score to all the candidate entities by using the cosine similarity between the document abstract $d$ and the summary of $e$'s Wikipedia article ($WS_e$).
\iffalse
, given by the formula $f_{1}(A, B)$ as:
\[
f_{1}(A,B) = \frac{\sum_{i=1}^{n} A_i B_i}{\sqrt{\sum_{i=1}^{n}{A_i}^2}{\sqrt{\sum_{i=1}^{n}{B_i}^2}}}
\]
where $A$ and $B$ are n-dimensional multinomial vectors of $d$ and $WS_{e}$ respectively. 
\fi
Based on these confidence scores, we prepare a ranked list of all the candidate Wikipedia articles for the given mention $m$. When the difference of the confidence scores between the highest and the second highest candidate thus produced is less than a given threshold ($TH_{cs}$), we mark this mapping to be ambiguous at this stage and use a metapath approach for further disambiguation; otherwise we choose the highest ranked candidate. In other words, a difference below the threshold indicates that we are not confident enough with the most agreeable candidate obtained from the cosine similarity method and therefore, resort to the metapath based approach. 
%needs to be resorted to.For the experiments, we have kept $TH_{cs}$ as 0.06.

\iffalse
We adopt a collective inference approach for linking the mentions $m$ to the maximum agreeable candidate $e$ if $|E| > 1$. For ranking each $e \: \epsilon \: E$, the collective inference approach takes the maximum intersection of possible n-grams of $e$'s summary page with the mentions set $M$ rather than analyzing each $m$ individually. %For example, given the sentence that contains the mentions $language \: modeling$ and $WSD$, the collective approach will analyze the two mentions simultaneously to determine the best candidate for $WSD$ as $Word \: Sense \: Disambiguation$ from the set of candidates $C$. %\todo{This is not clear at all. What is the collective inference approach?}

After applying collective inference, any tie among $c$'s is resolved by adopting a scoring function which calculates the cosine similarity between the document abstract $d$ and the summary of $m$'s Wikipedia page.
%, given by the formula $f_{1}(A, B)$ as:
\[
f_{1}(A,B) = \frac{\sum_{i=1}^{n} A_i B_i}{\sqrt{\sum_{i=1}^{n}{A_i}^2}{\sqrt{\sum_{i=1}^{n}{B_i}^2}}}
\]
where $A$ and $B$ are n-dimensional multinomial vectors of $d$ and $WS_{e}$ respectively. 
\fi
%\vspace{-2mm}

%\vspace{1mm}
%\todo{Start with a sentence or two on why metapaths are required for the ranking.}

%\todo{Also, you have not mentioned that your scoring function computes cosine similarity between abstracts and the first para of wiki page.}

%\vspace{-4mm}
\subsubsection{\bf{Candidate entity ranking using metapaths}}\label{candidate_rank}
As mentioned in Section~\ref{candidate_gen}, whenever the cosine similarity based approach is not very discriminative (less than $TH_{cs}$) for link disambiguation, we attempt to use a larger context for the particular entity mention. The concept of metapaths is used to identify documents related to the abstract $d$. These related documents are used to add more context in the process of choosing the most agreeable $e$ from the set $E$ for the given entity mention $m$. Next, we provide a brief overview of the concept of metapaths and how it can be used in the context of wikification of scientific documents.

\begin{figure}[!t]
    \centering
    \includegraphics[width=0.5 \textwidth, frame]{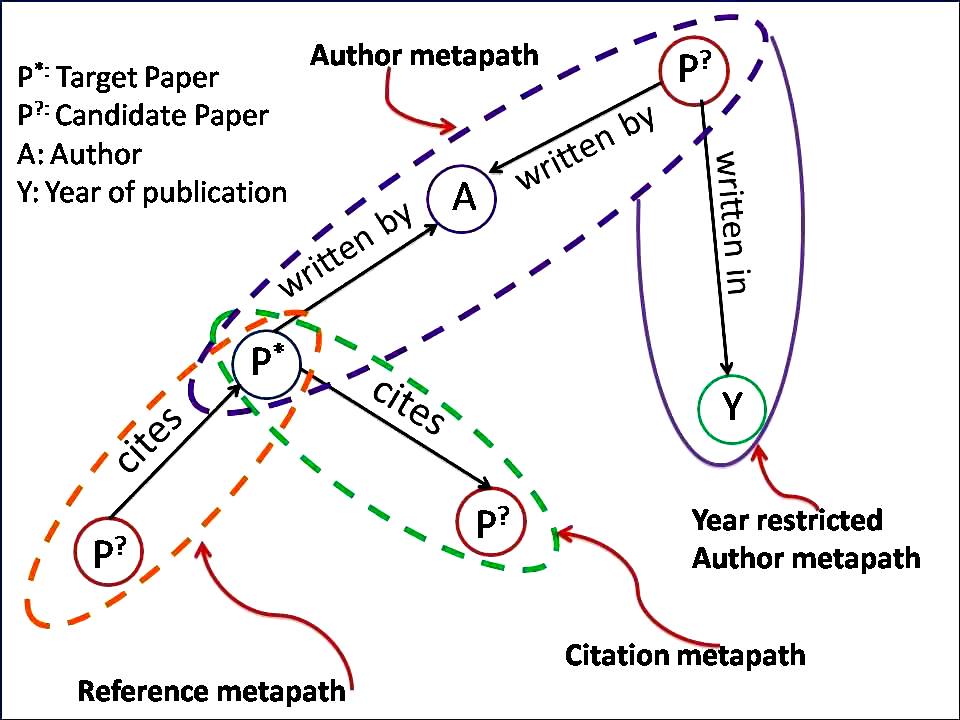}
    \caption{Metapaths in merged Citation and Author Publication Network. The network contains three types of objects: Paper (P), Author (A), Year (Y). }
    \label{fig:Metapaths in Citation Network}
 \vspace{-6mm}  
 %\vspace{1mm}
\end{figure}
%\vspace{1mm}
\iffalse
We choose to go for this metapath approach only if we are not confident enough with the most agreeable candidate obtained from the previous approach. Confidence is obtained by taking the difference in the highest and the second highest cosine similarity values among all the candidates $e \: \epsilon \: E$. A confidence score threshold, identified empirically as 0.06 is used.
\fi 
%\todo{Pawan: Again I think that this picture is very naive. The expressions later used should be perfect?}
%\todo{Replace co-author by author meta path in the figure}

A metapath is a path composed of sequence of relations between different object types defined over a heterogeneous network structure. We mainly focus on two types of heterogeneous network - citation network and author publication network. Citation network is a network where each node represents a scientific article and edges are of two types - `cites`, `cited by'. On the other hand, author publication network is a network where nodes are of types `author', `scientific article', `year of publication' etc. and edges are of types `written by', `written in' etc. In our experimental settings, we construct a network as shown in Figure~\ref{fig:Metapaths in Citation Network}, which is the merged version of both the citation network and the author publication network. We use the ACL Anthology Network-2013 dataset to prepare this network, which consists of more than 20000 paper (P) nodes, 17000 author (A) nodes and 40 year (Y) nodes. In this network on an average a paper gets 8 citations, and cites 6 papers; an author writes 3 papers on average, with maximum number of papers written by an author being 161; around 500 papers get published on average per year while the maximum number of papers published per year can go up to 1832.
We use this network to define three {\sl basic metapaths} - Citation (C), Reference (R) and Author (A), each of which represents a semantic relation, as follows:
%We then define the following three basic meta-paths:

%\begin{description}
\textbf{Author metapath (A)}:\begin{tikzcd}[row sep=0.4cm, column sep = 0.4cm]
P^{*} \arrow[r, "w"] & A & P^{?} \arrow{l}[swap]{w}
\end{tikzcd}

\textbf{Reference metapath (R)}:\begin{tikzcd}[row sep=0.4cm, column sep = 0.4cm]
P^{?} \arrow[r, "c"] & P^{*} 
\end{tikzcd}

\textbf{Citation metapath (C)}:\begin{tikzcd}[row sep=0.4cm, column sep = 0.4cm]
P^{*} \arrow[r, "c"] & P^{?} \\
\end{tikzcd}\\
%\end{description}
%\vspace{-2mm}
where $P^{*}$ is the target paper node in the graph and $P^{?}$ is the candidate same-author, cited or referenced paper node. 
For instance, author metapaths try to expand the document using other papers written by the same author $A$.

Further, we propose one {\sl restricted metapath} as follows, to help choosing only semantically related context- %We propose two restricted meta-paths as follows: %For instance, {\sl Keyword restricted metapath} indicates that the target paper's similar author / cited papers can be relevant for consideration, if these contribute to the relevant keyword, and {\sl Year-restricted metapath} indicates that the target paper's similar author papers can be relevant for consideration, if the paper is written within a backward window of five years.
%Each of these meta-paths represent one particular semantic relation. For instance, author meta-paths try to expand the document using other papers written by the same author. 
%We use ``restricted meta-paths" to enable us choosing only semantically related contexts. We propose two restricted meta-paths as follows:
%\[\begin{tikzcd}[row sep=0.4cm, column sep = 0.4cm]
%P^{*} \arrow[r, "w"] & A/C & P^{?} \arrow{l}[swap]{w} \arrow{r}{con} & K^{*}
%\end{tikzcd}\] which indicates that the target paper's similar author / cited papers can be relevant for consideration, if these contribute to the relevant keyword $K$, which is the mention under consideration, \todo{Abhik, is this correct?} and
%\vspace{-2mm}
\[\begin{tikzcd}
[row sep=0.4cm, column sep = 0.4cm]
P^{*} \arrow[r, "w"] & A & P^{?} \arrow{l}[swap]{w} \arrow{r}{in} & Y^{*}
%\vspace{-4mm}
\end{tikzcd}\]

which indicates that the target paper's  ($P^{*}$) similar author ($A$) papers are relevant for consideration only if the paper ($P^{?}$) is written within a backward window of some years ($Y$). %\todo{Do you consider only backward window? Yes, only backward window.}
Such restricted metapaths can be exploited to detect more semantically related contexts. In addition, we incorporate abstracts from the metapaths only if their cosine similarity with the target abstract is greater than a threshold $TH_{relevance}$.
We append the abstracts obtained using metapaths to the target abstract, thus enhancing its scientific context. We then provide a score to each $e \in E$; we take the maximum intersection of possible $n$-grams of this enhanced context of $d$ with $e$'s summary page, and %, rather than analyzing each mention $m$ individually. 
%The target abstract is expanded by appending it with the abstracts from the metapaths, if their cosine similarity with the target abstract is greater than 0.4. Then for ranking each $e \: \epsilon \: E$, the collective inference approach takes the maximum intersection of possible n-grams of $e$'s summary page enhanced with metapaths with the mentions set $M$ rather than analyzing each $m$ individually. %\todo{How can a summary page be enhanced with metapahts, which are defined for scientific abstracts?} 
the candidate with maximum intersection is taken as the correct disambiguated candidate entity for that mention $m$. %If there are more than one candidates with maximum intersection, we go with the first one. %\todo{Still not clear, where is e coming in picture? You are measuring similarity of the extended abstract (same for all entities) with the set of mentions in the abstract. That will give a single value, independent of the entity pages.}

\subsubsection{\bf{Further enhancements based on types of mentions}}
%\vspace{1mm}
On further analysis, we find that %the behavior of the algorithm is not consistent across all categories of mentions. More specifically, 
many mentions in scientific abstracts are acronyms. As per the Wikipedia style of disambiguation pages (\enquote{dab pages}), long disambiguation pages should be grouped into subject sections, and even subsections. These sections (and subsections) should typically be in alphabetical order, e.g.: `arts and entertainment'; `business'; `government and politics'; `places'; `science and technology'. In terms of organization of the Wikipedia dab pages, we find that acronyms generally tend to have long disambiguation pages. Some examples of acronyms are noted in Table~\ref{table:stats}. 
%For instance IR (for Information Retrieval) had 44 entries in its dab page with 5 sections and 2 subsection, and MT (for machine translation) had 83 entries with 11 sections and 4 subsections. 
We find that taking intersection from the extended abstract from metapaths sometimes leads to an inappropriate entity linking; therefore, a certain weight needs to be given to the original abstract. On the other hand, non-acronyms tend to have long dab pages (with sections and subsections) very rarely, and thus this problem does not arise.

\begin{table}
\begin{center}
%\resizebox{340.0pt}{!}{%
\begin{tabular}{ | M{4.1em}| M{2.6cm} | M{2.5cm} |} 
%\begin{tabular} {|c|c|c|c|c|}
\hline
\textbf{Sample Acronyms} & \textbf{Expansion} & \textbf{Number of dab entries and sections in dab page} \\ 
\hline
IR & Information Retrieval & 44 (5 sections \& 2 subsections) \\
\hline
MT & Machine Translation & 83 (11 sections \& 4 subsections)\\
\hline
TC & Text Categorization & 59 (12 sections)\\
\hline
DP & Detector of Presuppositions & 81 (10 sections)\\
\hline
\end{tabular}

\caption{Some representative acronym mentions with long dab pages.}
\label{table:stats}
\end{center}
\vspace{-10mm}
\end{table}

Thus we propose a separate approach to deal with the acronyms. After finding cosine similarity between the document abstract $d$ and the summary of $e$'s Wikipedia article, we incorporate abstracts from metapaths into a new scoring function $Score(n)$, which uses a linear interpolation for the cosine similarity of the original abstract and the expanded context using metapaths as follows:
%\vspace{-0.2cm}
\[Score(A,B,C) = 
	\begin{array} {lr}
	\alpha * f(A, B) + (1-\alpha) * f(C, B) \\
	\end{array}
 %   \vspace{-1mm}
\]

where A, B \& C are $n$-dimensional multinomial vectors of $d$, $WS_{e}$ and $(d\:+\:MP(d))$ respectively where $MP(d)$ = \\
$\{abstract\,(metapaths\,(d))\}$; $f()$ is the cosine similarity function, and $\alpha$ is a constant.
%empirically set to 0.6. %[{\color{red}{\bf AM: It is not clear why this scoring function is used.}}]
The top ranked candidate from the new scoring function $Score()$ is taken as the correct disambiguated link for that mention.  %Note that for acronyms, we are not using the collective inference mechanism. 
As described in Section~\ref{sec:experi}, for acronyms, going with this new scoring function \iffalse rather than the collective inference approach (which corresponds to ranking using n-gram intersection from the enhanced context only)\fi produces better results for majority of the cases. \iffalse Weight of 0.6 is given to the original abstract, as determined empirically.\fi %\todo{Abhik, is alpha 0.4 or 0.6?}
%\todo{This may not be sufficient motivation / explanation for taking different approaches.} 
%The pages in science and technology section normally gives higher $f_{2}(n)$ score than pages in other sections, thus leading to correct results.

 %The new scoring function after incorporating these metapaths uses a linear interpolation for the cosine similarity of the original abstract and the expanded context using metapaths (weight of 0.6 is given to the original abstract, as determined empirically). 
 %Remember, for non-acronyms, we are not using the scoring function $f_{2}(n)$. 
 
%\todo{Abhik, are you sure that for acronyms, cosine similarity was uses, while for non-acronyms, n-gram intersection was used?}

%\todo{Still the algorithm for ranking is not explained anywhere.}
%\vspace{-3mm}
\section{Experimental Results and Discussions}
\label{sec:experi}
We have broadly two phases in our approach - mention extraction and entity linking cum link disambiguation. First we conduct experiments to evaluate these two phases individually, and then we measure the full system's performance by considering both the phases together. We evaluate the performance of this system using the standard precision-recall metrics and compare it with various baselines. We compare the approach with mainly three baselines - TagMe \cite{ferragina2010tagme}, AIDA~\cite{yosef2011aida} and Wikifier~\cite{cheng2013relational}, among which TagMe turns out to be the most competitive baseline as per the evaluation results. We also tried Babelfy~\cite{Moro14entitylinking} but it disambiguates all the non-stop keywords (rather than just the relevant keywords). %it is not suitable for our experiments as  in the text as is clear from the Figure \ref{fig:babelfy} and \ref{fig:babelfy_ex}. So it is cumbersome to calculate precision-recall in our experimental settings. Further it uses BabelNet, which is a multilingual encyclopedic dictionary \& a semantic network, rather than simply leveraging on Wikipedia for linking entities. 
Thus we do not include the evaluation results from Babelfy.
%\todo{Abhik, if we mention the systems below and say that we are not comparing, we need to give a better justification. Please look into this.--}
As discussed previously, other existing systems,  like~\cite{guo2013link,huang2014collective} wikify microblog texts specifically tweets, whereas the baselines that we use in the experimental settings are able to wikify any type of short texts composed of few tens of terms, which makes them applicable for this purpose.
The evaluation results for each of the two phases, and comparisons with the baselines are described next. \\  

 %\todo{It is not clear as to how you aggregate the annotations from multiple annotators.}
%\todo{We need to provide statistics on the number of mentions per abstract in gold standard}
%\todo{sruthi/abhik: what is inter-annotator agreement?}

%TagMe can wikify any type of short texts composed of few tens of terms, which makes it applicable as a strong baseline, whereas other systems wikify any kind of standard texts or blog texts. AIDA is another web-based online interface which maps mentions of ambiguous names onto canonical entities like people or places, registered in a knowledge base like DBpedia, Freebase, or YAGO, given a natural-language text or a Web table. Wikifier is another tool for Wikification of texts by University of Illinois. Wikipedia Miner is another tool whose web service is not currently working and so we couldn't try out our experiments on it. 

%[{\color{red}{\bf AM: Rewrite the paragraph above adding all the baselines that you consider.}}]
%A mention and candidate pair $<m,\:c>$ is considered to be correct if and only if $m$ is linkable and $c$ is its correct candidate concept. 
%\todo{all: Is TagMe the most competitive baseline? If so, we should put up this point here. Otherwise we should justify why we have chosen this baseline?}\\
%\vspace{3mm}

\begin{table}[t!]
\begin{center}
%\resizebox{230.0pt}{!}{%
\begin{tabular}{ | M{5.2em}| M{5cm}|} 
%\begin{tabular} {|c|c|c|c|c|}
\hline
\textbf{Mention} & \textbf{Link of Wikipedia articles} \\ 
\hline
web server  & \url{https://en.wikipedia.org/Web_server} \\
\hline
MT &\url{https://en.wikipedia.org/wiki/Machine_translation} \\
\hline
natural language processing  
 &\url{https://en.wikipedia.org/wiki/Natural_language_processing} \\
\hline
WSD &\url{https://en.wikipedia.org/wiki/Word-sense_disambiguation} \\
\hline
\end{tabular}
\caption{Examples of annotated mentions in the gold standard dataset.}
\label{table:gs}
\end{center}
\vspace{-7mm}
\end{table}

\noindent {\em \textbf{Gold standard}}:
To evaluate the efficiency of our system WikiM, we prepare the gold-standard data from 2013 ACL Anthology Network (AAN) dataset. 50 random scientific article abstracts are taken from AAN dataset. These 50 abstracts are given to each of the 15 annotators with computational linguistics background, where they are asked to find out the terms or phrases from the abstracts to wikify and then link those terms or phrases with a Wikipedia article, without posing any limit to the number of abstracts they can wikify. On an average each annotator annotated 17 abstracts. Table \ref{table:gs} presents some of these annotated mentions from the gold standard dataset.
%, whereas maximum 38 abstracts and minimum 6 abstracts were annotated by some of the annotators.

%Since all the annotators are from Computational Linguistic background, only the relevant scientific keywords are chosen by them. 

\begin{table}[t!]
\begin{center}
%\resizebox{220.0pt}{!}{%
\begin{tabular}{ | M{5em} | M{1.7cm} | M{1.8cm}| M{1.8cm} |} 
%\begin{tabular} {|c|c|c|c|c|}
\hline
\textbf{Abstract Category} & \textbf{\# abstracts in the dataset} & \textbf{Avg \# single-word mentions (Gold standard/WikiM)} & \textbf{Avg \# Multi-word mentions (Gold standard/WikiM)} \\ 
\hline
4-mention abstract & 6 & 3.16/2.83 & 1.83/1.17 \\
\hline
8-mention abstract & 25 & 5.72/4.5 & 2.72/3.5 \\
\hline 
12-mention abstract & 19 & 7.05/8.67 & 2.58/3.33 \\ 
\hline
\end{tabular}

\caption{Statistics of single-word and multi-word mentions in the gold standard dataset and in the result of WikiM}

\label{table:sing-mult}
\end{center}
\vspace{-8mm}
\end{table}
%\vspace{10mm}

%\todo{Abhik, did all 15 wikify each abstract, or was it distributed?}
Then we use the union\footnote{Taking intersection to aggregate, leads to very few (2-3) mentions per document.} of annotations from multiple annotators in order to aggregate. These manually annotated 50 random scientific article abstracts constitute our gold standard dataset. The full gold standard dataset contains an average of 8.5 mentions per abstract. Table \ref{table:sing-mult} gives the statistics of both the single and multi-word mentions in the dataset \footnote{In the statistics acronyms are considered to be single-word mentions}. In this setting, where each abstract can be wikified by any number of annotators, each of whom can provide any number of mentions, computing the inter-annotator agreement is not relevant. 
We use 10 random abstracts from this gold standard dataset as the validation dataset for entity linking phase to set the parameter ($TH_{cs}$, $TH_{relevance}$, $\alpha$) values. %which gives best performance for these 10 abstracts . 
The sensitivity analysis of these parameters and the corresponding performance measures are reported later. We compare the performance of WikiM along with all the baselines against the rest 40 abstracts from the gold standard dataset. Note that we have also tuned the baselines' parameter values wherever it is possible using validation dataset. For example, TagMe has a parameter $\epsilon$, which can be used to fine tune the disambiguation process. As per the TagMe documentation, a higher value favors the most-common topics (e.g., in tweets, where context may not be much reliable), whereas a lower value uses more context information for disambiguation. TagMe gives the best performance for $\epsilon=0.2$.\\

%with several values of epsilon(hyperparameter used to tune TagMe) while using TagMe API for our experiments and found that for epsilon=0.3, TagMe is giving the best performance. In this paper all the results of TagMe reported are for the same value(0.3) of the hyperparameter(epsilon). \\
%We test WikiM on the rest 40 abstracts from gold standard dataset and report the performance in the following sections along with several baselines' performance.\\

\noindent {\em \textbf{Mention extraction}}: 
%We first measure the precision-recall of the mention extraction phase, followed by analysis of entity linking precision.   
%To evaluate the performance of this wikification system, we use the standard precision, recall and F-measures. 
Statistics from the evaluation of mention extraction phase are presented in Table~\ref{table:1}. %[{\color{red}{\bf Add results from other baselines in Table \ref{table:1}.}}] 
We see that our mention extraction approach boosts the precision and recall of the system from the range of 8.54\%-30.56\% to 72.42\% and 2.42\%-39.03\% to 72.1\% respectively. Even though our mention extraction approach is very simple, the reason for better results of mention extraction phase could be that we compute idf from the corpus of scientific abstracts only. This clearly shows the advantage of using a system, specifically built for scientific articles. 
%\todo{one sentence may be deleted.}
%[{\color{red}{\bf AM: Put results from all the competing baselines.}}] 
Keyphrase or mention extraction snapshots of all the baselines for a representative abstract are given in Figure \ref{fig:me_baselines}.
%\footnote{Banerjee, Sujata, et al. \enquote{Scalable grid service discovery based on uddi.} Proceedings of the 3rd international workshop on middleware for grid computing. ACM, 2005.} 
%\todo{Can we remove Babelfy, and put the remaining three in one row?} 
%Here we have given an abstract of a scientific article as input to all the baselines including Babelfy and systems' outputs are presented in Figure \ref{fig:me_baselines}. Note that Babelfy considers all the non-stopwords as possible mentions. 
It shows that AIDA mainly chooses acronyms as the mentions to wikify, which is true for Wikifier as well, whereas TagMe also chooses other potential mentions besides acronyms. The results from WikiM are shown in Figure \ref{fig:our_algo}. We see that our approach chooses more appropriate mentions, e.g., `scalable', `grid computing', `deployment', `UDDI' etc., compared to other baselines including both acronyms and non-acronyms.  
We also test and compare our algorithm on benchmark dataset used in the SemEval-2010 Task-5 on keyword extraction from scientific articles~\cite{Kim:2010:STA:1859664.1859668} with other state-of-the-art baselines. Even though we are using simple tf-idf based ranking approach for mention extraction, we see comparable performance in terms of precision and recall with other baselines. As the main focus of our work is entity linking, we do not explore further to improve the mention extraction algorithm which gives an otherwise decent performance.   \\  

%\todo{can you highlight some important differences using these figures?}

%\subsection{Experimental Results}

\begin{table}[t!]
\begin{center}
%\resizebox{230.0pt}{!}{%
\begin{tabular}{ | M{1.35cm} | M{1.9cm}| M{1.cm} | M{1.8cm} |} 
%\begin{tabular} {|c|c|c|c|c|}
\hline
\textbf{Method} & \textbf{Precision} & \textbf{Recall} & \textbf{F-Measure} \\ 
\hline
AIDA & 8.54\% & 2.42\% & 3.62\% \\
\hline
Wikifier & 23.88\% & 5.76\% & 8.31\% \\
\hline 
TagMe & 30.56\% & 39.03\% & 32.65\% \\ 
\hline
WikiM & \bf72.42\% & \bf72.1\% & \bf71.52\% \\ 
\hline
\end{tabular}

\caption{Evaluation of Mention Extraction w.r.t. AAN 2013 dataset.}
\label{table:1}
\end{center}
\vspace{-7mm}
\end{table}
%\vspace{3mm}

%\noindent {\em \textbf{Comparison based on SemEval2010 Task-5}}:\\
%To further strengthen the evaluation of mention extraction phase, we test our algorithm on benchmark dataset used in the SemEval-2010 Task-5 on keyword extraction from scientific articles~\cite{Kim:2010:STA:1859664.1859668}. %We compare our mention extraction approach with other state-of-the-art baselines on this dataset. 
%Task-5 of the Semantic Evaluation 2010 (SemEval-2010) workshop~\cite{Kim:2010:STA:1859664.1859668} was to The task concerned automatic extraction of key-phrases or keywords from scientific articles. 
%It was a shared task for keyphrase extraction, aimed at providing a standard assessment to benchmark current approaches. For this task, the committee compiled a set of 284 scientific articles from the ACM Digital Library, with keyphrases carefully chosen by both their authors and readers. Table~\ref{semeval} shows the performance of our system and all other baselines over this SemEval2010 dataset. %[{\color{red}{\bf AM: Why not all other possible baselines?}}] 
%The best F-measure reported was 27.50\%, achieved by the HUMB team. It is to be noted from the table that by using the same experimental setup as that of SemEval2010 Task-5 for mention extraction, we outperform all other baselines in terms of F-measure. We are achieving comparable results in terms of precision and recall. %\todo{Update based on new results.}  
%hat our system outperforms the then submitted systems. 
\begin{figure*}[t!]
\centering
\subfigure[AIDA]{
    \includegraphics[width=0.475 \textwidth, height=0.25 \textheight, frame]{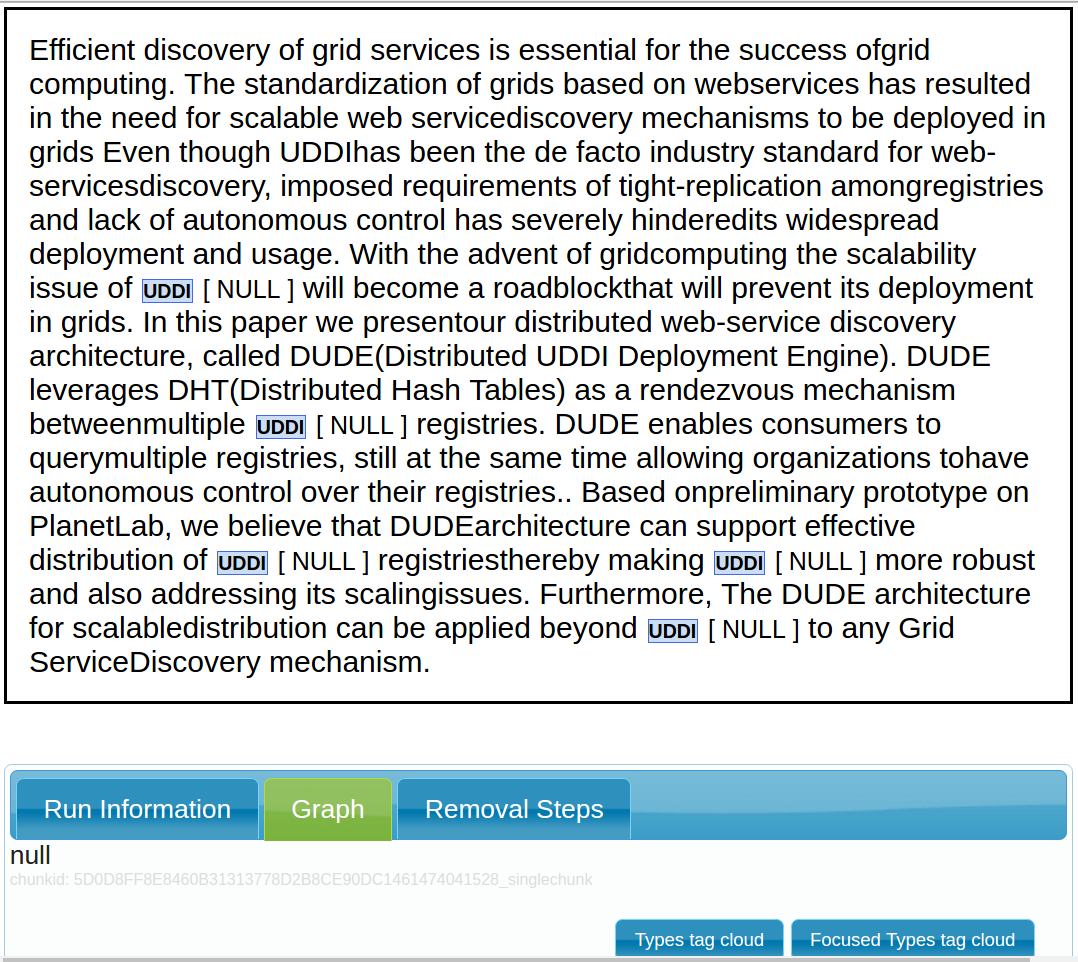}
    \label{fig:AIDA}
}
\subfigure[Wikifier]{
    \includegraphics[width=0.475 \textwidth, height=0.25 \textheight, frame]{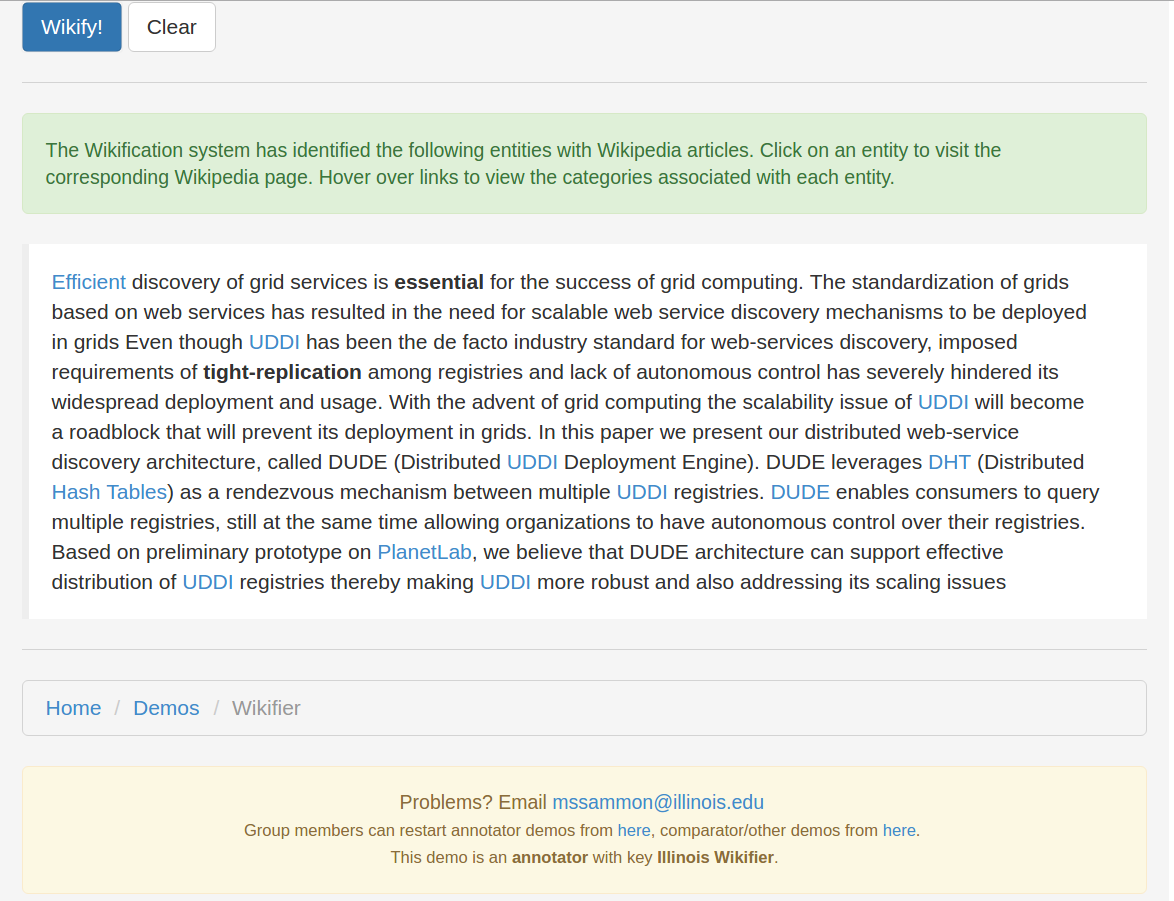}       
    \label{fig:wikifier}
}
\subfigure[TagMe]{%\cite{ferragina2010tagme}
    \includegraphics[width=0.475 \textwidth, height=0.25 \textheight, frame]{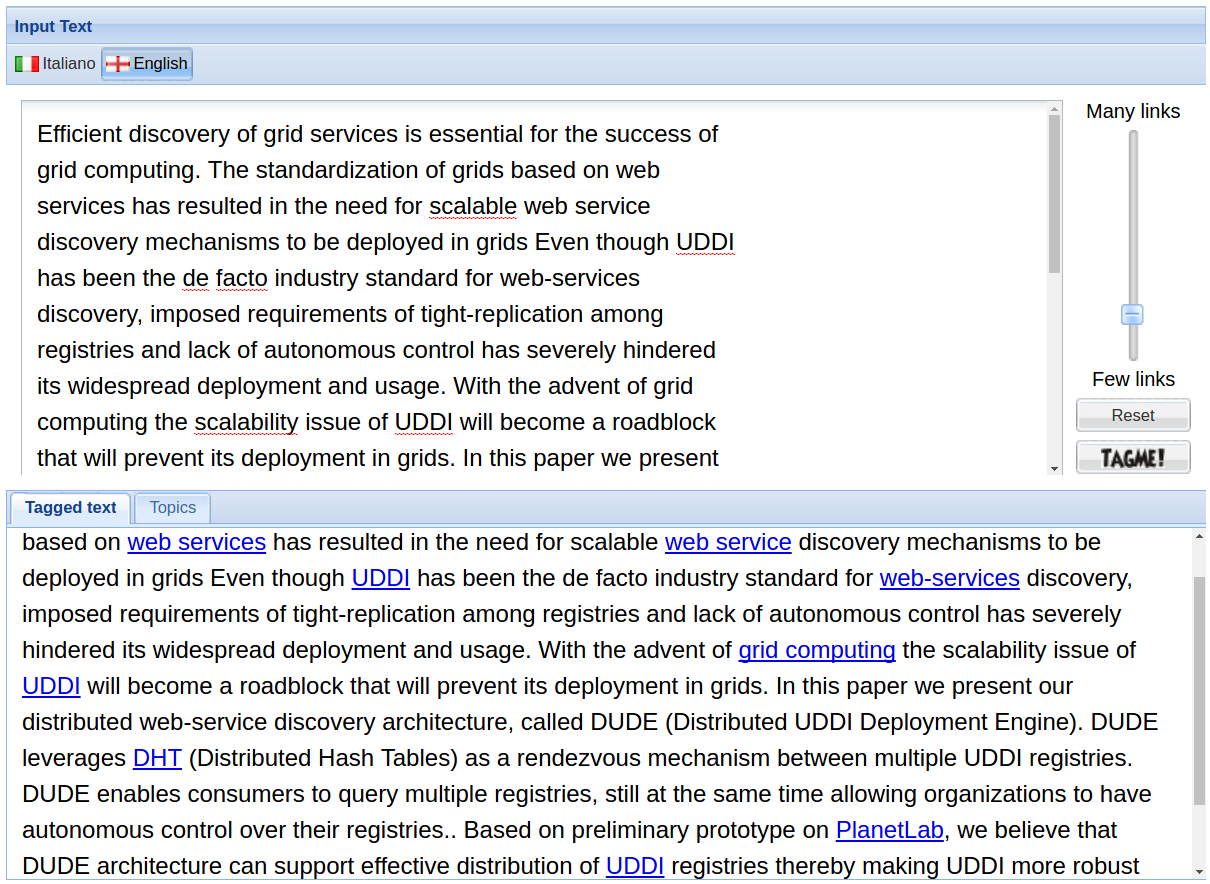}    
    \label{fig:tagme}
}
\subfigure[WikiM]{%\cite{Moro14entitylinking}]{
 \includegraphics[width=0.475 \textwidth, height=0.25 \textheight]{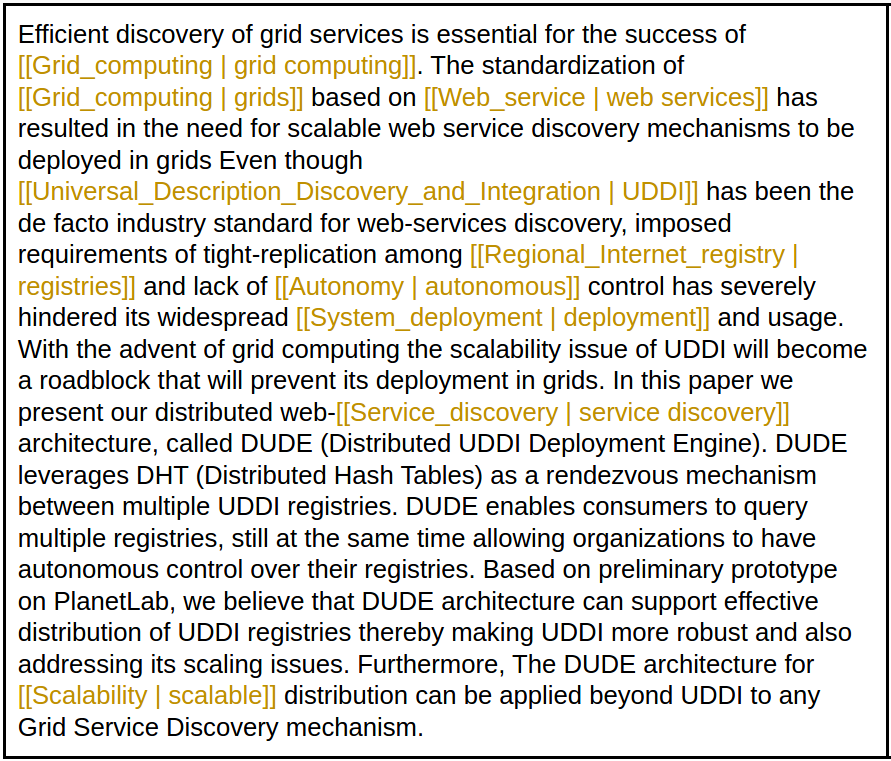}
    %\vspace{-4mm}
    \label{fig:our_algo}
    }
\vspace{-4mm}
\caption[]{Mention extraction by other baselines and WikiM.}
\label{fig:me_baselines}
%\vspace{-3mm}
\end{figure*}

\if{0}
\begin{figure*}[t!]
\centering
\subfigure[AIDA]{
    \includegraphics[width=0.31 \textwidth, height=0.2 \textheight, frame]{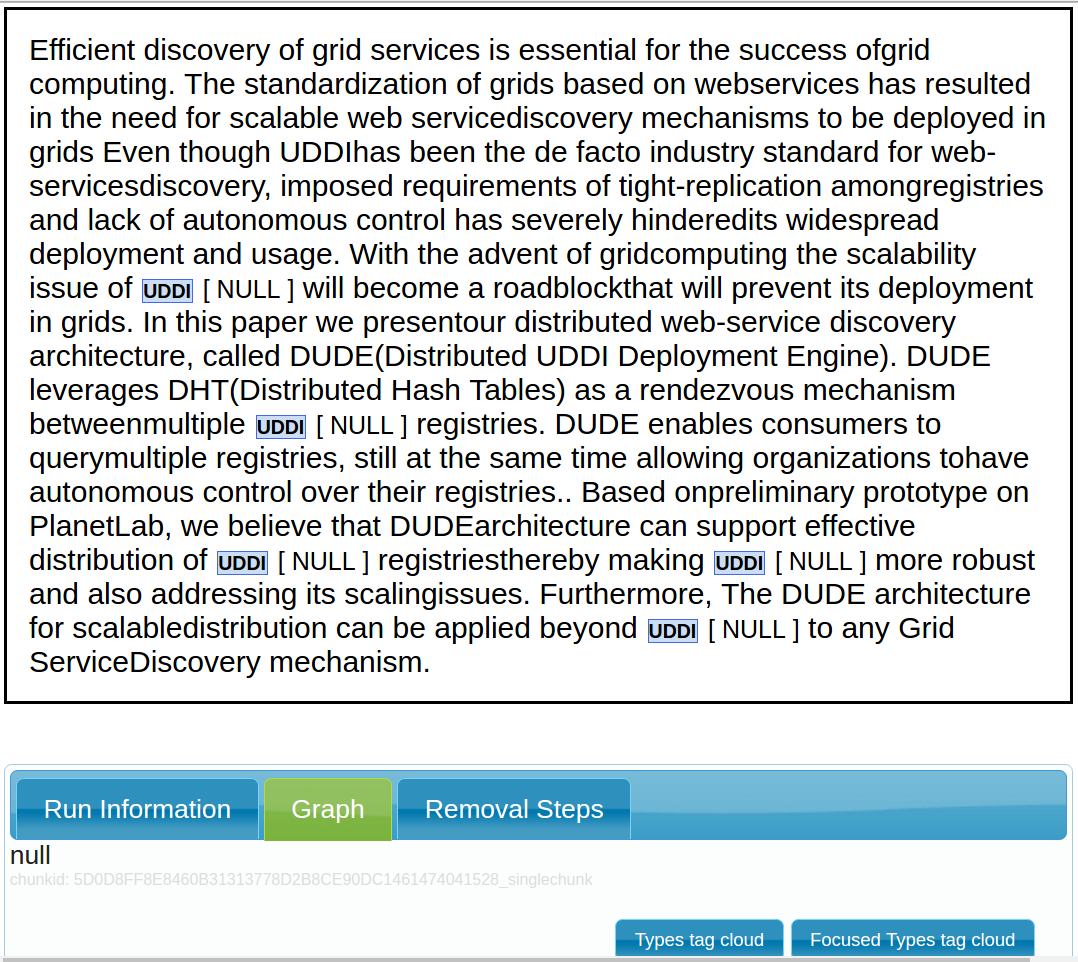}
    \label{fig:AIDA}
}
\subfigure[Wikifier]{
    \includegraphics[width=0.31 \textwidth, height=0.2 \textheight, frame]{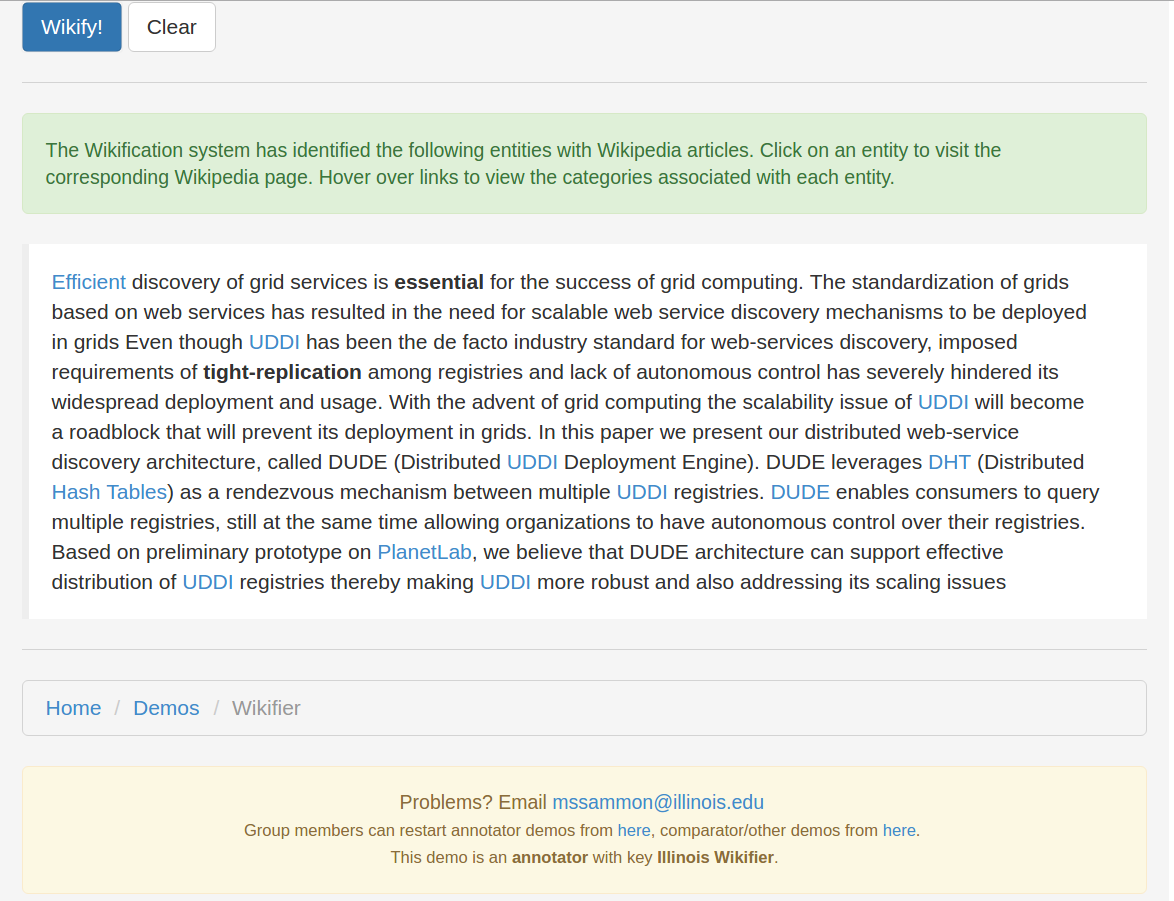}       
    \label{fig:wikifier}
}
%\subfigure[Babelfy, from Universidad Carlos III de Madrid]{
%    \includegraphics[width=0.475 \textwidth, height=0.2 \textheight, frame]{babelfy.png}    
%    \label{fig:babelfy}
%}
\subfigure[TagMe]{
    \includegraphics[width=0.31 \textwidth, height=0.2 \textheight, frame]{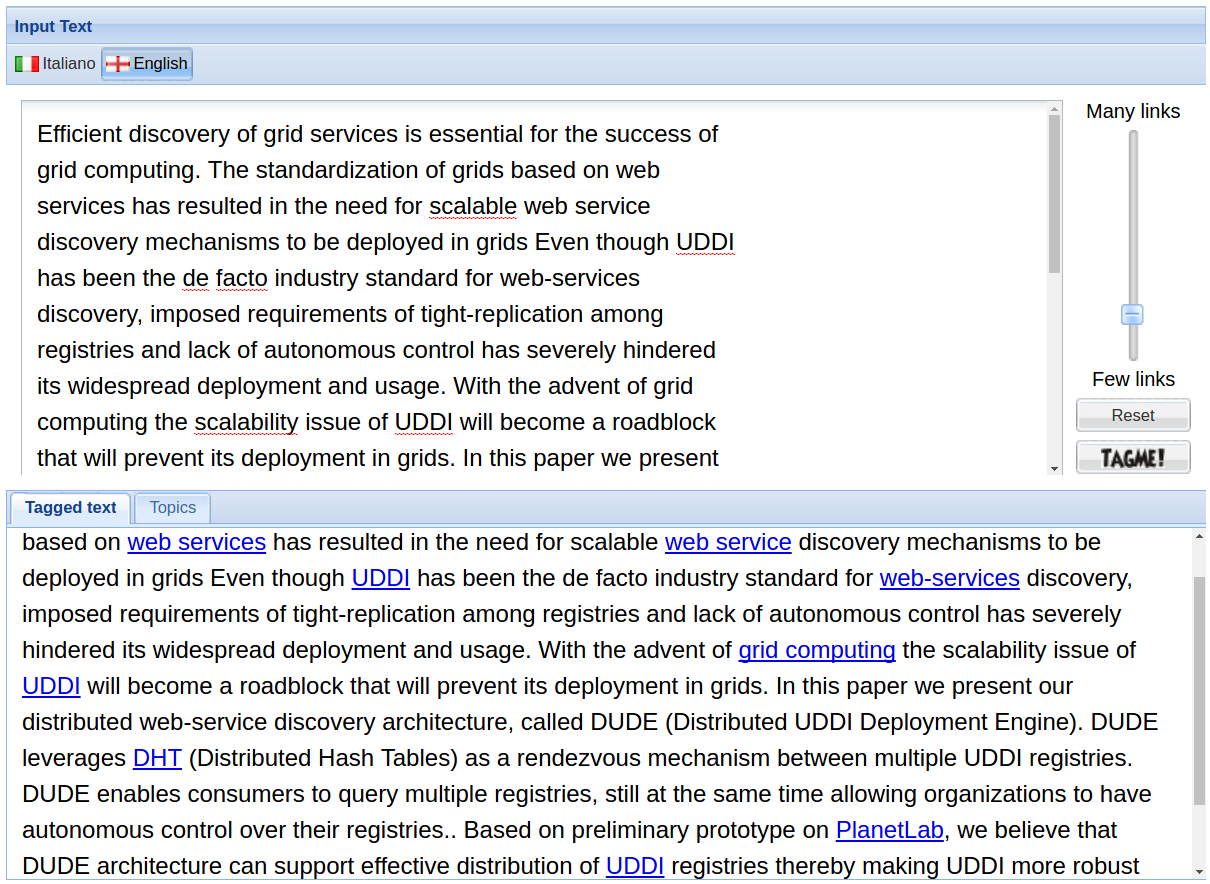}    
    \label{fig:tagme}
}
\vspace{-4mm}
\caption[]{Mention Extraction by other baselines and WikiM}
\label{fig:me_baselines}
\vspace{-4mm}
\end{figure*}
\fi
\if{0}
\begin{figure}[t!]
    \centering
    \includegraphics[width=0.475 \textwidth, height=0.2 \textheight]{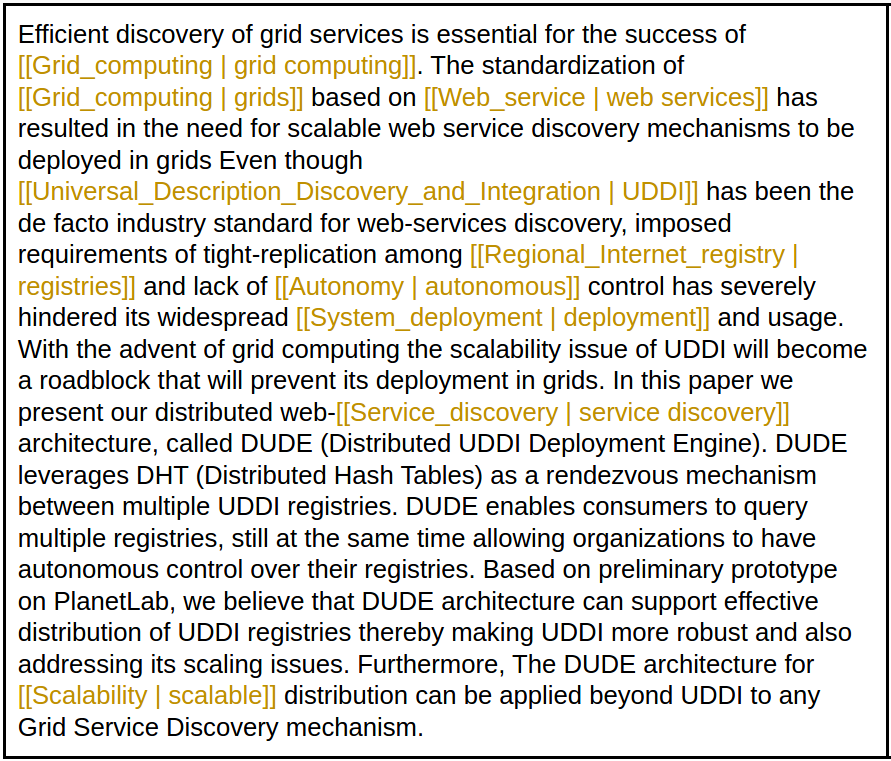}
    %\vspace{-4mm}
    \caption{Mention Extraction by WikiM.}
    \label{fig:our_algo}
 %   \vspace{-2mm}
\end{figure}
\fi
%\todo{The next line is not clear. Each system will give a selection. ou need to compare this selection with the gold standard.}
%In order to compute the correctness of mention extraction we 
%we have considered only those terms or phrases which have a valid wikipedia page to link with. 

%Note that in this phase of mention extraction only the mentions (a single- or multi-word expression) having a positive linking  validity are considered for processing. 
%\todo{But where is the meta-path coming into picture in these values? It is not clear.}
%compares the performances of Keyword Extraction phase of the devised Wikification system with respect to $TagMe$.

%In this work, only the fragments (a single- or multi-word expression) having a positive linking  validity is considered for processing. Since all the annotators were made aware of this fact and were from Computational Linguistic background, only the most relevant scientific keywords were chosen for this task. Table \ref{table:1} compares the performances of Keyword Extraction phase of the devised Wikification system with respect to $TagMe$. %Keyword Precision implies the fraction of retrieved keywords that are relevant whereas Keyword Recall implies the fraction of relevant keywords that are retrieved by the system, both with respect to the gold standard. 

\begin{table}[t!]
\begin{center}
%\resizebox{230.0pt}{!}{%
\begin{tabular}{ | M{13.5em} | M{2.6cm}|} 
%\begin{tabular} {|c|c|c|c|c|}
\hline
\textbf{Method} & \textbf{Link Precision} \\ 
\hline 
AIDA & 6\% \\
\hline
Wikifier  & 19.33\% \\
\hline
TagMe & 58.91\% \\ 
\hline
WikiM - CR metapaths & 69.41\% \\ 
\hline
WikiM - Author metapaths & 71.4\% \\ 
\hline
WikiM - CRA metapaths & \bf73\% \\ 
\hline
WikiM - Year restricted CRA metapaths & \bf73.80\% \\ 
\hline
\end{tabular}

\caption{Evaluation of Entity Linking w.r.t. AAN 2013 data set.}
\label{table:2}
\end{center}
\vspace{-10mm}
\end{table}
%\vspace{1mm}

\noindent{\em \textbf{Entity linking}}: The comparative evaluation of entity linking is shown in Table~\ref{table:2}, evaluated only on those mentions, that are adjudged to be extracted correctly by the system because of the availability of gold standard. % is available only for those mentions. 
%Now for evaluating the entity linking performance, we have considered only those mentions which are correctly extracted from the abstracts by our system. 
% represents the comparative evaluation results of entity linking for the baseline as well as various combinations of meta-paths. 
%We compare the results of wikification of scientific document abstracts with various baselines as shown in Table~\ref{table:2}. 
The result showcases the link precision (only for true positives) 
%and overall recall 
for all these baselines, of which TagMe is again found to be the most competitive one. Note that our system gives significant improvement over TagMe for link precision as well.
%Note that we have also tried with several values of epsilon(hyperparameter used to tune TagMe) while using TagMe API for our experiments and found that for epsilon=0.3, TagMe is giving the best performance. In this paper all the results of TagMe reported are for the same value(0.3) of the hyperparameter(epsilon). 
%. 
%marginal, the significant improvement in mention extraction phase along with non-negligible improvement in linking phase makes WikiM a more accurate and reliable system than TagMe.   
%\todo{Abhik, are you sure TagMe performs so poor for link precision?}

\begin{figure}[t!]
\centering
%\subfigure[For the mention 'Tagging']{
    \includegraphics[width=0.475 \textwidth, height=0.25 \textheight, frame]{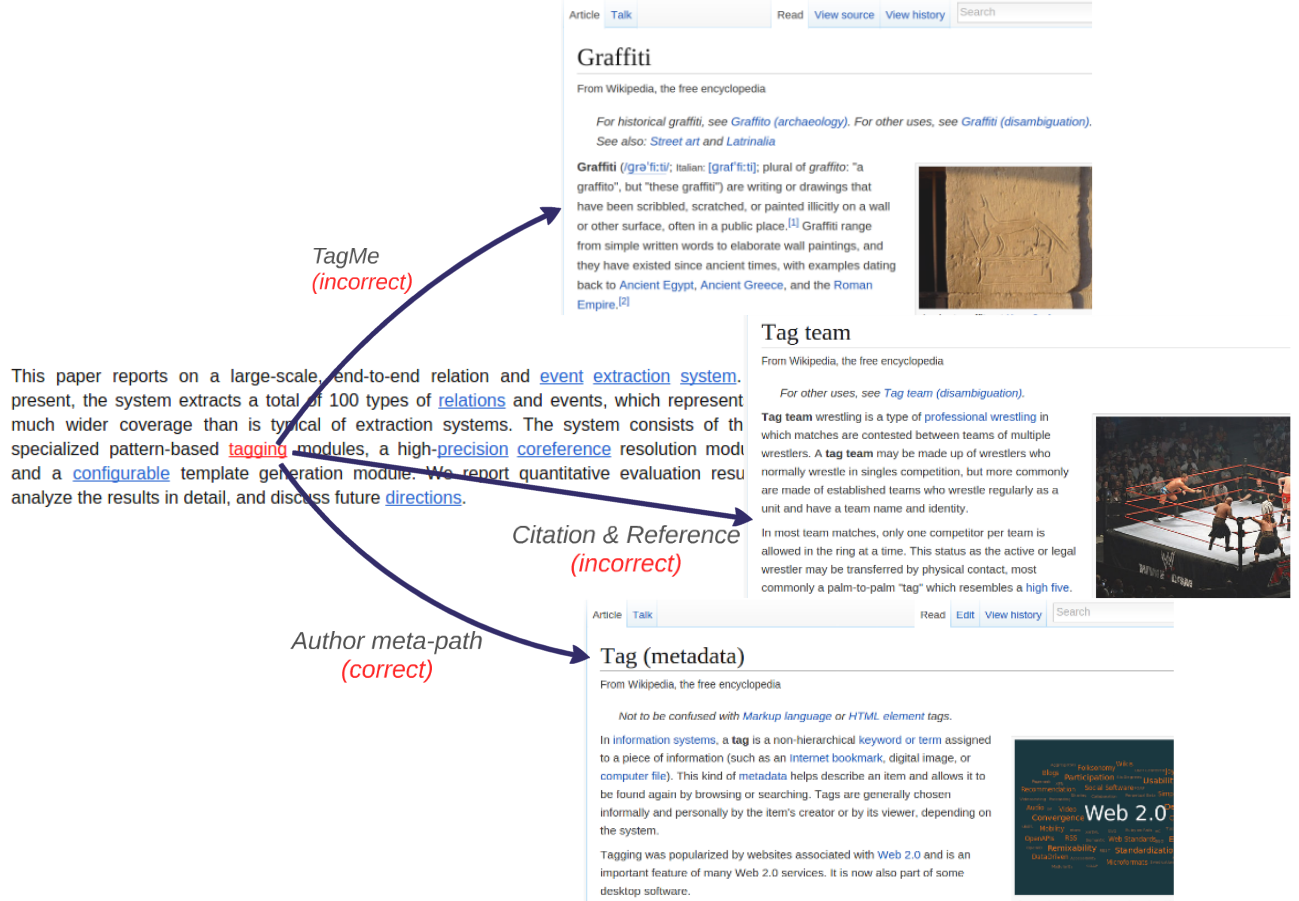}
%    \label{fig:tagging}
%}
%\subfigure[For the mention 'Transcriptions']{
%    \includegraphics[width=0.475 \textwidth, height=0.2 %\textheight, frame]{trans}       
%    \label{fig:trans}
%}
%\vspace{-2mm}
\caption[]{Links given by various metapaths for the mention `Tagging'.}
\label{fig:metapath_var}
%\vspace{-5mm}
\end{figure}

The relatively low performance of the baselines demonstrates that relying on prior popularity within a small document abstract alone may not suffice for the wikification of scientific documents. For instance, it is difficult to link the mention `WSD' from a small abstract, to `Word Sense Disambiguation' without much contextual information. Our approach uses the metapath information from citation and author publication networks, which adds to the context of the seed document. Thus, the chances of linking the mention `WSD' to the most popular concept `World Sousveillance Day' by mistake are by far reduced. We can see that in comparison to all the baselines, our method using author metapaths relying on all the paper abstracts written by the same author as that of the seed paper, achieves comparable performance. For example as shown in Figure~\ref{fig:metapath_var}, the mention `tagging' in a scientific point of view is correctly linked to `tag (metadata)' since other papers written by the same author share similar contexts. We can also see that incorporating citation and reference metapaths into author metapaths provides further gains, indicating the effectiveness of enhanced context using these metapaths. We achieve marginal improvement while using year restricted CRA metapath (with a back-window of 5 years), over the baselines, showing that incorporating related abstracts in the context coupled with timeliness is beneficial as well.
\begin{table*}[!tbh]
\begin{center}
%\resizebox{230.0pt}{!}{%
%\begin{tabular}{ | M{8.8em}| M{2.8cm} | M{4.2cm}|} 
\begin{tabular} {| M{4em}| M{1.5cm} | M{7cm}|M{6cm} |}
\hline
\textbf{Baseline} &\textbf{Mention} & \textbf{Link of the gold standard Wikipedia article}&\textbf{Link provided by baseline}  \\ 
\hline
 & NLP & \url{https://en.wikipedia.org/wiki/Natural_language_processing} & \url{https://en.wikipedia.org/wiki/Nonlinear_programming} \\ \cline{2-4}
 AIDA & internet & \url{https://en.wikipedia.org/wiki/Internet} & \url{https://en.wikipedia.org/wiki/Youtube} \\
\hline
& NLG & \url{https://en.wikipedia.org/wiki/Natural_language_generation} & \url{https://en.wikipedia.org/wiki/Dutch_guilder} \\ \cline{2-4}
 Wikifier & ELS & \url{https://en.wikipedia.org/wiki/ELS_Language_Centers} & \url{https://en.wikipedia.org/wiki/London_Buses_route_ELW
} \\
\hline
& IR & \url{https://en.wikipedia.org/wiki/Information_retrieval} & \url{https://en.wikipedia.org/wiki/Infrared} \\ \cline{2-4}
 TagMe & integrate & \url{https://en.wikipedia.org/wiki/System_integration} & \url{https://en.wikipedia.org/wiki/Globalization
} \\
\hline
\end{tabular}

\caption{Sample cases where baselines provide erroneous entity linking.}
\label{table:baselines_error}
\end{center}
\vspace{-7mm}
\end{table*}

Cases where the baselines are linking to a wrong entity page, are shown in Table \ref{table:baselines_error}. It shows that, in almost all the cases the baselines fail to utilize the context where the mention exists. The probable reason of WikiM's better performance could be taking into account more contexts by incorporating the abstracts from different metapaths in order to comprehend the context of the mention.  
%\todo{The improvement over TagMe is marginal. Do we have a defense for this?}

%over the baselines (a minimum of 8.88\% link precision gain), showing that incorporating related abstracts in the context coupled with timeliness is highly beneficial. \todo{Is it still valid? I see that year-restricted metapaths are doing slightly worse in Table 3.} For instance, the mention `transcription' is correctly linked to 'transcriptions (software)' by using year restricted CRA metapaths while it has been linked to `transcription (linguistics)' by incorporating citation and reference information of similarly devised software to the author metapaths, and to `transcription (service)' with just the earlier method. \todo{This example is not self-explanatory. How would the referee know which transcription is correct?}  %However our approach fails to link some mentions like `alignment' to their correct links, hence making it open for further improvements. 

\begin{figure}[t!]
\centering
%\subfigure[Result of Babelfy]{
%    \includegraphics[width=0.475 \textwidth, height=0.2 \textheight, frame]{babelfy_ex}
%    \label{fig:babelfy_ex}
%}
%\subfigure[Result of TagMe]{
   \includegraphics[width=0.475 \textwidth, height=0.25 \textheight, frame]{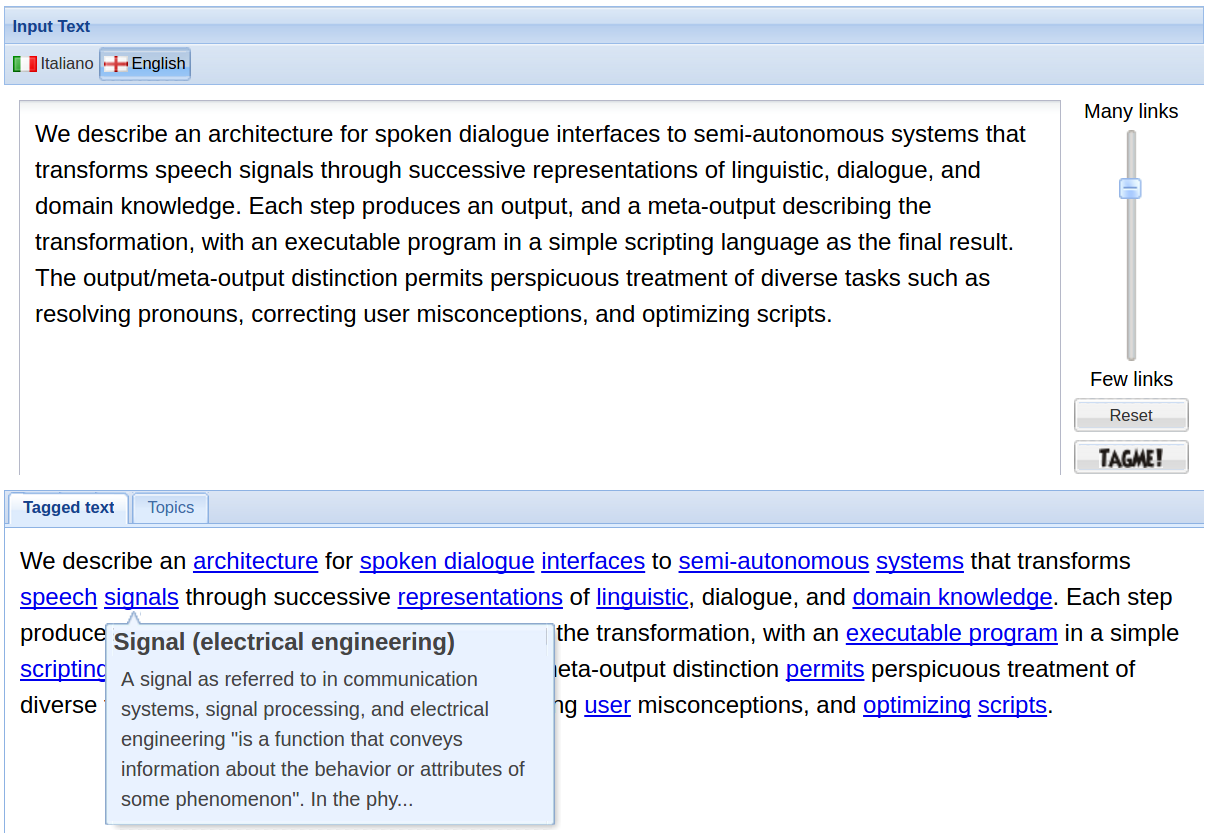}       
    %\label{fig:tagme_ex}
%}
\vspace{-4mm}
\caption[]{Linking mention `Signal' by TagMe.  %\todo{Why not other baselines here}
}
\label{fig:le_baselines}
\vspace{-1mm}
\end{figure}

\begin{figure}[t!]
    \centering
    \includegraphics[width=0.475 \textwidth]{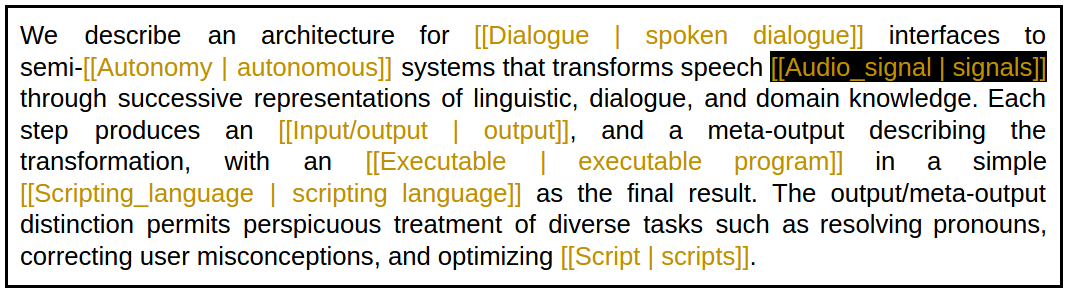}
\vspace{-5mm}
\caption{Linking mention `Signal' by WikiM.}
    \label{fig:my_algo_ex}
    \vspace{-1mm}
\end{figure}

Links given for the mention `signal' (a representative example) by TagMe and WikiM, are shown in Figure~\ref{fig:le_baselines} and Figure~\ref{fig:my_algo_ex} respectively. The correct meaning of the mention `signal' in the context of the given abstract is `signal related to speech or audio'. Since Wikipedia is having quite a few dab page entries for the mention `signal', it is very tricky to choose between them. However, from the snapshots given here, it can be well adjudged that our approach works better than TagMe. TagMe provides `signal (electrical engineering)' as the correct link while there are more appropriate entries in Wikipedia. As Figure \ref{fig:my_algo_ex} shows, WikiM produces the exact meaningful link with respect to the context of the current document. Other baselines such as AIDA and Wikifier do not provide an annotation for the mention `signal' for this representative example, and therefore are not shown here. \\
%\todo{You can remove Babelfy from here, since we are not comparing with it anyway.}

%\begin{table}[t!]
%\begin{center}
%\resizebox{200.0pt}{!}{%
%\begin{tabular}{ | M{17.0em} | M{2.5cm}|} 
%\begin{tabular} {|c|c|c|}
%\hline
%\textbf{Baselines} & \textbf{Link Precision (True Positives)} \\ 
%\hline
%Wikifier Demo - University of Illinois & 61.79\% \\
%\hline 
%AIDA - Accurate Online Disambiguation of NEs in Texts \& Tables & 65.42\% \\
%\hline
%TagMe & 84.20\% \\ 
%\hline 
%Wikipedia Miner & \multicolumn{1}{M{\combinedlength}|}{Web service not working} \\
%\hline 
%Babelfy & \multicolumn{1}{M{\combinedlength}|}{Disambiguates all the non-stop words (rather than relevant keywords alone), so is tricky to calculate precision-recall.} \\
%\hline
%Our approach & 91.71\% \\
%\hline
%\end{tabular}
%}
%\caption{Comparison with baselines}
%\label{baselines}
%\end{center}
%\end{table}

\begin{table}[t!]
\begin{center}
%\resizebox{230.0pt}{!}{%
\begin{tabular}{ | M{10.8em} | M{3.5cm}|} 
%\begin{tabular} {|c|c|c|c|c|}
\hline
\textbf{Algorithm} & \textbf{Link Precision} \\ 
\hline
Same Algorithm for both & 63.92\% \\
\hline
Separate algorithms for both & \bf67.08\% \\ 
\hline
\end{tabular}

\caption{(Validation set) Effect of acronyms \& non-acronyms distinction on entity linking.}
\label{acr}
\end{center}
\vspace{-9mm}
\end{table}

%\todo{May be, we can present these results before the comparisons with baselines.}
\noindent{\em \textbf{Analysis of the effect of the parameters:}} We attempt to analyze the importance of various parameters used in our approach using a validation set of 10 random abstracts from the gold standard dataset, which we also use to tune the parameters. First, we analyze the effect of using \textbf{separate algorithms for acronyms and non-acronyms}. This affects only the link precision. For the AAN dataset, 17 out of 50 document abstracts contained acronyms as mentions in the gold standard and 30\% of those acronyms do not have their full forms in the abstract. As shown in Table~\ref{acr}, using separate algorithms for acronyms and non-acronyms gives a link precision of 67.08\%, as compared to 63.92\% obtained using the same algorithm for both while tested on the validation set.
%This made us to analyze the differences in structure of Wikipedia pages for most of the acronyms and non-acronyms, and propose separate algorithm accordingly. 
%For both of these algorithms, the preceding candidate generation phase is the same and requires a confidence score (empirically determined as 0.06) as discussed in section~\ref{candidate_gen}. 
Next, to analyze the effect of the \textbf{$TH_{cs}$} (difference between the cosine similarity score of top two entities), we conduct experiments with and without keeping $TH_{cs}$, along with various values for this threshold on the validation set. The results are presented in Table~\ref{conf}. 
%compares the results with and without the confidence score. 
The results show that while the performance is not very sensitive to $TH_{cs}$, it gives a small improvement and a value of $0.06$ gives the best performance. If we do not keep the threshold $TH_{cs}$, which means that irrespective of the difference of the top two candidate Wikipages' confidence score, metapath based approach would be taken for further processing, the link precision drops by a small margin. We also see that the link precision drops significantly to 63.3\% when metapaths are not considered at all.
%\todo{What does it mean to not have this threshold? Does it mean that we always choose the top-ranked mention and never take the metapath branch? If so, then as the performance is not sensitive to keeping this threshold, one can easily argue that metapaths are not very useful in enhancing performance -- this might be very self-defeating. However, if not keeping the threshold means you always take the metapath branch no matter what is the difference in scores, then we have stronger evidence that metapath is useful.} %confirms that confidence score is one of the key parameters which helps to boost the performance of the system.
We can see the similar effect for the variation of parameter $TH_{relevance}$ in Table~\ref{relevance}. It shows that the results are quite sensitive, and a value of $0.4$ yields the best performance on the validation set.

Similarly, to see the effect of the parameter $\alpha$ in the scoring function for acronyms, we experiment with various values of $\alpha$ as noted in Table~\ref{alpha}. Our system gives the best performance for the $\alpha$ value of 0.6. 

Using this validation dataset, we set the parameters of the system ($TH_{cs}$, $TH_{relevance}$,$\alpha$) to (0.06,0.4,0.6) for all the experiments.\\

\begin{table}[t!]
\begin{center}
\resizebox{230.0pt}{!}{%
\begin{tabular}{ | M{12.8em} | M{4cm}|} 
%\begin{tabular} {|c|c|c|c|c|}
\hline
\textbf{Value of $TH_{cs}$} & \textbf{Link Precision} \\ 
\hline
without threshold (always consider metapath) & 66\% \\
\hline 
0.02 & 59\% \\
\hline
0.04 & 62\% \\
\hline
0.06 & \bf67.08\% \\
\hline
0.08 & 59\% \\
\hline
never consider metapath & 63.33\% \\
\hline
\end{tabular}
}
\caption{(Validation set) Effect of variation of parameter $TH_{cs}$ on link precision.}
\label{conf}
\end{center}
\vspace{-7mm}
\end{table}

\begin{table}[t!]
\begin{center}
\resizebox{230.0pt}{!}{%
\begin{tabular}{ | M{12.8em} | M{4cm}|} 
%\begin{tabular} {|c|c|c|c|c|}
\hline
\textbf{Value of $TH_{relevance}$} & \textbf{Link Precision} \\ 
\hline
\bf{0} (take all the abstracts from metapath) & 64.4\% \\
\hline
0.2 & 57.5\% \\
\hline
\bf0.4 & \bf67.08\% \\
\hline
0.6 & 60.6\% \\
\hline
0.8 & 60.4\% \\
\hline
\end{tabular}
}
\caption{(Validation set) Effect of variation of parameter $TH_{relevance}$ on link precision.}
\label{relevance}
\end{center}
\vspace{-6mm}
\end{table}

\begin{table}[t!]
\begin{center}
%\resizebox{230.0pt}{!}{%
\begin{tabular}{ | M{8.8em} | M{4cm}|} 
%\begin{tabular} {|c|c|c|c|c|}
\hline
\textbf{Values for $\alpha$} & \textbf{Link Precision} \\ 
\hline
$\alpha$ = 0.5 & 64\% \\
\hline
$\alpha$ = 0.6 & \bf67.08\% \\
\hline
$\alpha$ = 0.7 & 63\% \\ 
\hline
\end{tabular}
%}
\caption{(Validation set) Effect of variation of parameter $\alpha$ on link precision.}
\label{alpha}
\end{center}
\vspace{-4mm}
\end{table}
%\vspace{3mm}
%Table \ref{table:3} demonstrates  This demonstration is done by dividing zones into three, on the basis of meta-path counts. This in turn indicates that link precision largely varies with diversity in context. CRA with restricted meta-path results in more or less consistent values as compared to more diverse CRA meta-path. 
\begin{table}[t!]
\begin{center}
\resizebox{230.0pt}{!}{%
\begin{tabular}{ | M{10.5em} | M{1.1cm}| M{1.1cm} | M{1.1cm} |} 
%\begin{tabular} {|c|c|c|c|c|}
\hline
\textbf{Method} & \textbf{Low ($<5$)} & \textbf{Med ($\ge 5, <10$)} & \textbf{High ($>10$)} \\ 
\hline
CRA Metapaths & 73.07\% & 64.96\% & 75.95\% \\
\hline
Year restricted CRA Metapaths & 71.75\% & 67.81\% & 86.67\% \\ 
\hline
\end{tabular}
}
\caption{Studying the effect of number of citations: citation-zone based evaluation (link precision) of entity linking. The citation-zones are indicated in the parenthesis.}
\label{table:3}
\end{center}
\vspace{-4mm}
\end{table}
%\vspace{3mm}
\begin{table}[t!]
\begin{center}
\resizebox{230.0pt}{!}{%
\begin{tabular}{ | M{12.8em} | M{2.6cm}|} 
%\begin{tabular} {|c|c|c|c|c|}
\hline
\textbf{Method} & \textbf{Overall Recall} \\ 
\hline
AIDA & 1\% \\
\hline
Wikifier & 4.1\% \\
\hline
TagMe & 25.76 \% \\
\hline
WikiM - CRA Metapaths & \bf52.75\% \\ 
\hline
\end{tabular}
}
\caption{Comparison of WikiM with all the baselines for the full system recall}
\label{table:4}
\end{center}
\vspace{-10mm}
\end{table}
%\vspace{1mm}
%\vspace{3mm}
\begin{table*}[!tbh]
\begin{center}
%\resizebox{230.0pt}{!}{%
%\begin{tabular}{ | M{8.8em}| M{2.8cm} | M{4.2cm}|} 
\begin{tabular} {| M{5.5em}| M{3cm} | M{2.8cm}|M{3cm} | M{1.6cm}|M{2.5cm} |}
\hline
\textbf{Method} &\textbf{Precision (Mention Extraction)} & \textbf{Recall (Mention Extraction)}&\textbf{F-Measure (Mention Extraction)} & \textbf{Link Precision}&\textbf{Full System Recall} \\ 
\hline
TagMe & 18\% & 42\% & 23\% & 54\% & 35\%  \\
\hline
WikiM & 26\% & 68\% & 34\% & 64\% & 50\% \\
\hline
\end{tabular}

\caption{Comparison of WikiM with TagMe while taking individual single annotators as ground truth. Mean values have been reported over 15 annotators.}
\label{table:Sar}
\end{center}
\vspace{-7mm}
\end{table*}

\begin{table}[t!]
\begin{center}
%\resizebox{230.0pt}{!}{%
\begin{tabular}{ | M{3.3em}| M{3cm} | M{3cm}|} 
%\begin{tabular} {|c|c|c|c|c|}
\hline
\textbf{Mention} & \textbf{Link of Wikipedia article with same surface form} & \textbf{Link of disambiguation page} \\ 
\hline
Java & \url{https://en.wikipedia.org /wiki/Java} \bf[an island of Indonesia] & \url{https://en.wikipedia.org/wiki/Java_(disambiguation)}\\
\hline
Tree &\url{https://en.wikipedia.org /wiki/Tree} \bf[a perennial plant with an elongated stem] & \url{https://en.wikipedia.org/wiki/Tree_(disambiguation)} \\
\hline
\end{tabular}

\caption{Example mentions having a disambiguation page but the library returns another page (erroneous) with the same surface form.}
\label{table:table_both}
\end{center}
\vspace{-6mm}
\end{table}

\begin{table}[t!]
\begin{center}
%\resizebox{230.0pt}{!}{%
\begin{tabular}{ | M{3.3em}| M{3cm} | M{3cm}|} 
%\begin{tabular} {|c|c|c|c|c|}
\hline
\textbf{Mention} & \textbf{Link of gold standard Wikipedia article} & \textbf{Link of wikipedia article by WikiM} \\ 
\hline
modify & \url{https://en.wikipedia.org /wiki/Editing} & \url{https://en.wikipedia.org/wiki/Modding}\\
%\hline
%effort &\url{https://en.wikipedia.org /wiki/Manual_labour} & \url{https://en.wikipedia.org/wiki/Effort,_Pennsylvania} \\
\hline
precise &\url{https://en.wikipedia.org /wiki/Accuracy_and_precision} & \url{https://en.wikipedia.org/wiki/Promise} \\
\hline

\end{tabular}

\caption{Example cases where gold standard Wikipedia article is not in the set of articles to be disambiguated returned by `wikipedia' library}
\label{table:table_n}
\end{center}
\vspace{-6mm}
\end{table}

\begin{table}[t!]
\begin{center}
%\resizebox{230.0pt}{!}{%
\begin{tabular}{ | M{3.3em}| M{2.2cm} | M{2.2cm}| M{1.5cm} |} 
%\begin{tabular} {|c|c|c|c|c|}
\hline
\textbf{Mention} & \textbf{Link of gold standard wikipedia article} & \textbf{Link of wikipedia article by WikiM} & \textbf{Number of abstracts in the metapath}  \\ 
\hline
schemes & \url{https://en.wikipedia.org /wiki/Scheme_(linguistics)} & \url{https://en.wikipedia.org/wiki/Scheme_(Mathematics)} & 72\\
\hline
transcrip-tions & \url{https://en.wikipedia.org/wiki/Orthographic_(transcriptions)} & \url{https://en.wikipedia.org/wiki/Transcription_(linguistics)} & 38 \\
\hline
romanian &\url{https://en.wikipedia.org /wiki/Romanian_language} & \url{https://en.wikipedia.org/wiki/Romania} & 33 \\
\hline

\end{tabular}

\caption{Example cases where too many abstracts from metapaths lead to wrong entity linking.}
\label{table:table_ne}
\end{center}
\vspace{-6mm}
\end{table}

\noindent{\em \textbf{Performance dependence on metapath counts}}: We further analyze the effectiveness of varied proportion of added metapaths. We divide the papers into low, medium and high zones, based on the count of citations. The results presented in Table \ref{table:3} imply that link precision for the papers from high zone and mid zone increases if we use year restricted methapaths as it reduces the diversity of context. But year restricted CRA metapath approach does not help improve the link precision of low zoned papers as the number of contexts from the metapath are already very few in this case.\\
%even lesser.\\  
%\todo{This is no more valid from the results. Can you please rewrite this if the results are updated?}
%\vspace{3mm}

\noindent {\em \textbf{Full system recall}}: A mention and a candidate pair $<m,\:e>$ is considered to be correct if and only if $m$ is linkable and $e$ is its correct candidate concept. Following this definition of correctness, we measure the recall of the full system. %We have compared in Table \ref{table:4}. 
%\todo{all: how is this recall defined; it may not be apparent. Add a footnote to define this.} 
The results in Table~\ref{table:4} confirm that our system's performance is significantly better than all the baselines in terms of overall effectiveness, outperforming the most competing baseline TagMe by a significant margin. \\
%\todo{Give imp. in percentage}. %[{\color{red}{\bf AM: Here again, put results from all possible competing baselines.}}]  
%vspace{-2mm}
%\vspace{3mm}
\begin{table}[t!]
\begin{center}
%\resizebox{220.0pt}{!}{%
\begin{tabular}{ | M{8.8em}| M{2.8cm} | M{1.5cm}|} 
%\begin{tabular} {|c|c|c|c|c|}
\hline
\textbf{Method} & \textbf{Mention Extraction Precision} & \textbf{Link Precision} \\ 
\hline
Majority Decision & 94.03\% & 73.23\% \\
\hline
Macro-Averaging & 89.36\% & 71.11\% \\
\hline
Micro-Averaging & 88.81\% & 69.53\%\\
\hline
\end{tabular}
\caption{Evaluation results for Top Cited articles from AAN 2013 data set.}
\label{table:micro-macro}
\end{center}
\vspace{-6mm}
\end{table}
%\fi
%\if{0}
\begin{table}[t!]
\begin{center}
%\resizebox{230.0pt}{!}{%
\begin{tabular}{ | M{8.8em}| M{2.8cm} | M{1.5cm}|} 
%\begin{tabular} {|c|c|c|c|c|}
\hline
\textbf{Method} & \textbf{Mention Extraction Precision} & \textbf{Link Precision} \\ 
\hline
Majority Decision & 67.09\% & 67.67\% \\
\hline
Macro-Averaging & 65.98\% & 71.8\% \\
\hline
Micro-Averaging & 63.41\% & 67.71\%\\
\hline
\end{tabular}

\caption{Evaluation results for data mining articles from MAS data set.}
\label{table:MAS}
\end{center}
\vspace{-6mm}
\end{table}
%\vspace{3mm}
\begin{table}[t!]
\begin{center}
%\resizebox{230.0pt}{!}{%
\begin{tabular}{ | M{8.8em}| M{2.8cm} | M{1.5cm}|} 
%\begin{tabular} {|c|c|c|c|c|}
\hline
\textbf{Method} & \textbf{Mention Extraction Precision} & \textbf{Link Precision} \\ 
\hline
Majority Decision & 86.6\% & 73.29\% \\
\hline
Macro-Averaging & 82.37\% & 69.55\% \\
\hline
Micro-Averaging & 85.04\% & 69.33\%\\
\hline
\end{tabular}

\caption{Evaluation results for bio-medical data set.}
\label{table:biomed}
\end{center}
\vspace{-10mm}
\end{table}

\noindent {\em \textbf{Single annotator result}}:
In order to verify that the comparison results are not due to some bias in our experimental settings, we compare the performance of WikiM and the most competitive baseline TagMe on single annotator results as well. The average measures computed by taking each individual annotator's annotations as ground truth are given in Table~\ref{table:Sar}. We also see that out of 15 annotators, WikiM does better than TagMe in $12,11,12,9,10$ cases for the measures presented in Table~\ref{table:Sar} respectively. The consistency of WikiM's good performance in this experiment shows that the gold standard creation based on the union of the annotations from 15 annotators does not make the evaluation procedure biased.\\

\noindent {\em \textbf{Error analysis}}: Even though we achieve significant performance boost over the baselines, we investigate further to point out the erroneous cases for which WikiM links the mention to a wrong Wikipedia entity page and try to find out the possible reasons for the same.  
As discussed earlier, in order to collect all the Wikipedia entities with surface form similar to the mention's surface form, we have used python library `wikipedia' which wraps the MediaWiki API. For this purpose, we use wikipedia.page() function which takes as argument a word and returns either the Wikipedia article with the same surface form as the word or a list of Wikipedia articles present in that word's disambiguation page. Nevertheless, this function has some limitations which leads to poor performance for some cases. Some example cases are shown in Table~\ref{table:table_both}, where for a particular mention both the  disambiguation page as well as the page having the same surface form exist. The function in the wikipedia library returns only the page with the same surface form for a given mention, causing no candidate entities to be disambiguated, which %are having both Wikipedia articles with the same surface form and the disambiguation pages. In these cases the library function however does not return the entries in the disambiguation page. So the entity linking phase of WikiM does not receive the set of Wikipedia articles to disambiguate and this 
leads to linking those mentions wrongly. There are also some other cases, where even though the library function returns a set of Wikipedia articles to disambiguate for a given mention, the gold standard article is not in that set and hence ends in a wrong entity linking. Some of the examples are shown in Table~\ref{table:table_n}. In future, we plan to get rid of these errors by pre-processing the wikidump ourselves instead of using the MediaWiki API.  

Further, while we see that for most of the cases extending the target abstract by taking relevant abstracts from the metapath helps in link disambiguation, there are a few cases where this broadening of context leads to wrong links. Table~\ref{table:table_ne} shows some of those examples. The probable reason could be that in all these cases the number of relevant abstracts from the metapath is too large to be informative. The possible solution could be to put another constraint in terms of the importance of a paper while adding abstracts from the metapaths along with textual relevance. The importance of an article could be measured by its citation count, PageRank in the citation graph etc. Note that, this step of incorporating importance to improve the performance of WikiM would however lead to the introduction of one more parameter to the system. \\

\noindent {\em \textbf{Evaluation of top cited documents from AAN 2013 data set}}: To investigate if having many metapaths has any adverse effect on the performance of the system, a dataset consisting of the top 50 cited articles (should usually have a large number of metapaths) is taken from AAN 2013 dataset. Table~\ref{table:micro-macro} gives the keyword precision and link precision (for true positives alone) values for the top-50 cited documents from the this dataset. The results of WikiM for each of these document abstracts are evaluated by three annotators independently. 
%\todo{8 or 3?} 
For each wikified mention in an abstract, the annotators are asked to choose among the following three options: i). keyword correct and link correct, ii). keyword correct but link incorrect, iii). keyword incorrect. Thus, the link precision is computed only if the annotator responded using one of the first two options. We report the keyword precision and link precision based on three evaluation criteria: \textbf{majority decision} is taken from the agreement of at least two out of the three annotators. 
%\todo{update this.} 
\textbf{Macro-averaged precision} is calculated by first taking the average precision of each abstract (which in turn is calculated as the fraction of annotators who agree to a particular case statement for each keyword, averaged over all the keywords in that abstract) and then taking the average of these. \textbf{Micro-averaged precision} results are calculated by computing precision for each wikified mention (fraction of annotators agreeing) and taking an average of all the mentions in all the 50 abstracts. %constructing contingency table (with precision values for all the keywords), and then calculating precision using these sums. 
We see that even in this dataset, link precision achieved by WikiM is around 70\% for all the three evaluation criteria. %While the keyword precision is 89.36\% as per the majority decision, it is always above 70\% using any of the evaluation criteria. 
The keyword precision is close to 90\% using any of the evaluation criteria. Thus, the performance of this system is quite consistent even on the dataset with many metapaths. \\
%\todo{update}
%\vspace{3mm}

\noindent {\em \textbf{Evaluation of data mining documents from MAS dataset}}: To further verify the consistency of our approach, we test its performance on a different dataset of scientific articles. 50 scientific abstracts are taken from Microsoft Academic Search (MAS) dataset\footnote{https://academic.microsoft.com/} in the domain of `data mining' and are wikified using WikiM. The evaluation framework is similar to the one used for the top cited articles from AAN dataset. Each of these abstracts is evaluated by three annotators independently.  %\todo{Did you use 8 anotators again?} The results by our algorithm were verified by 8 annotators. 
%In a similar way here also we have prepared the gold-standard dataset by manual evaluation. 
The results presented in Table~\ref{table:MAS} show that even when we move on to the domain of `data mining', the performance of the system is quite consistent. The link precision is at least 67\% using any of the evaluation criteria which is consistent with the evaluation results of the gold standard data. As per the majority decision, the keyword precision is 67.67\%. 
%our system is performing consistently well for both the tasks of mention extraction and entity linking.     
Since we use Wikipedia as the knowledge base, the links provided by WikiM confine to the existing Wikipedia entries. There are cases where all the currently existing candidate pages in Wikipedia are inappropriate for the mention with respect to the context of the document under observation. For instance, mentions such as `splits' from the AAN data set and `scientific data' from data mining domain of MAS data set, do not have valid Wikipedia pages for their corresponding meaning; thus, the links proposed by WikiM are labeled as incorrect by the annotators.\\

\noindent {\em \textbf{Evaluation of bio-medical dataset}}: We further test our system on a completely different dataset of scientific articles. 50 scientific article abstracts taken from the bio-medical dataset\footnote{It contains more than 1.1 million Bio-medical articles, downloaded from
\url{http://www.ncbi.nlm.nih.gov/pmc/tools/ftp/\#XML_for_Data_Mining}} are wikified using WikiM. %\todo{Where is the description of this dataset?} 
According to the results presented in Table~\ref{table:biomed}, we see that moving into completely different domain does not affect the consistency of WikiM much. 
\section{Conclusion}
%\vspace{-2mm}
This paper addresses a novel wikification approach for scientific articles. We use a tf-idf based approach to find out important terms to wikify. Then 
%we apply collective inference approach which analyze relations among multiple mentions and rank the candidate links. 
we facilitate the step of ranking the candidate links with metapaths extracted from citation and author publication networks of scientific articles. Experimental results show that the proposed approach helps to significantly improve the performance of the wikification task on scientific articles. 
%For the task of mention extraction, the performance of the system is found to be competitive on a benchmark SemEval dataset. 
The performance of our system is tested across various datasets, and the results are found to be consistent.
%The executable and annotated dataset used in our experiments are available at an anonymous link\footnote{\url{http://tinyurl.com/zm7vxq5}}. 
We plan to do similar experiments on a larger population and more variety of abstracts in future to further strengthen the necessity of wikification. Immediate future work should focus on some additional methods to detect the mentions more precisely, which may lead to overall improved entire system's performance. Our future plan is to incorporate the metapath based approach for mention extraction phase as well, and study the scope of metapath based joint mention extraction and entity linking system for this purpose.

%\bibliographystyle{ACM-Reference-Format}
%\bibliography{sigproc} 

\end{document}